\documentclass{mnras}

\usepackage{float}
\usepackage{tikz}
\usepackage{graphicx}
\usepackage{times}
\usepackage{mathtools}
\usepackage{amsmath}
\usepackage{amssymb}
\usepackage{verbatim}
\usepackage{bm}

\usepackage{yfonts}

\pdfoutput=1

\catcode`\@=11
\newcommand\gsim{\ifmmode{\mathrel{\mathpalette\@versim>}}
    \else{$\mathrel{\mathpalette\@versim>}$}\fi}
\newcommand\lsim{\ifmmode{\mathrel{\mathpalette\@versim<}}
    \else{$\mathrel{\mathpalette\@versim<}$}\fi}
\catcode`\@=12

\DeclareMathAlphabet{\mathpzc}{OT1}{pzc}{m}{it}

\newcommand\xv{{\bf x}}

\newcommand\Sigmas{\Sigma_*}

\newcommand\Sign{\Sigma_{\rm n}}

\newcommand\rhotil{\tilde\rho}
\newcommand\rtz{\tilde{\rho}_0}
\newcommand\rtu{\tilde{\rho}_1}
\newcommand\rtd{\tilde{\rho}_2}

\newcommand\rhos{\rho_*}
\newcommand\rtsz{\tilde{\rho}_{*0}}
\newcommand\rtsu{\tilde{\rho}_{*1}}
\newcommand\rtsd{\tilde{\rho}_{*2}}
\newcommand\rtsi{\tilde{\rho}_{*i}}

\newcommand\rhog{\rho_{\rm g}}
\newcommand\trhog{\tilde{\rho}_{\rm g}}
\newcommand\rtgz{\tilde{\rho}_{\rm g0}}
\newcommand\rtgu{\tilde{\rho}_{\rm g1}}
\newcommand\rtgd{\tilde{\rho}_{\rm g2}}
\newcommand\rtgi{\tilde{\rho}_{{\rm g}i}}

\newcommand\vrtz{\tilde{\varrho}_0}
\newcommand\vrtu{\tilde{\varrho}_1}
\newcommand\vrtd{\tilde{\varrho}_2}

\newcommand\rhoDM{\rho_{\rm DM}}
\newcommand\rhon{\rho_{\rm n}}

\newcommand\ms{m_*}
\newcommand\Es{E_*}
\newcommand\mg{m_{\rm g}}
\newcommand\Eg{E_{\rm g}}
\newcommand\Hf{{\cal F}}
\newcommand\I{{\cal I}}
\newcommand\Ic{{\cal I}_{\rm c}}
\newcommand\Ieq{{\cal I}_{\frac{\upi}{2}}}
\newcommand\Iinf{{\cal I}_{\infty}}
\newcommand\Izero{{\cal I}_0}
\newcommand\intI{{\rm int}\hspace{0.2mm}({\cal I})}

\newcommand\qs{q_*}
\newcommand\qg{q_{\rm g}}
\newcommand\etas{\eta_*}
\newcommand\etag{\eta_{\rm g}}
\newcommand{\vcirc}{v_{\rm c}}
\newcommand{\Og}{\Omega_{\rm g}}
\newcommand{\Os}{\Omega_*}

\newcommand\Rtil{\tilde{R}}
\newcommand\ztil{\tilde{z}}

\newcommand\rs{r_*}
\newcommand\rg{r_{\rm g}}
\newcommand\reff{R_{\rm e}}

\newcommand\Psis{\Psi_*}
\newcommand\PsiT{\Psi_{\rm T}}
\newcommand\Psibh{\Psi_{\rm BH}}
\newcommand\Psidm{\Psi_{\rm DM}}
\newcommand\Psig{\Psi_{\rm g}}
\newcommand\Psin{\Psi_{\rm n}}

\newcommand\tPsig{\tilde{\Psi}_{\rm g}}
\newcommand\tPsigz{\tilde{\Psi}_{\rm g0}}
\newcommand\tPsigu{\tilde{\Psi}_{\rm g1}}
\newcommand\tPsigd{\tilde{\Psi}_{\rm g2}}
\newcommand\tPsigi{\tilde{\Psi}_{{\rm g}i}}

\newcommand\tPsiz{\tilde{\Psi}_0}
\newcommand\tPsiu{\tilde{\Psi}_1}
\newcommand\tPsid{\tilde{\Psi}_2}

\newcommand\tpsiz{\tilde{\psi}_0}
\newcommand\tpsiu{\tilde{\psi}_1}
\newcommand\tpsid{\tilde{\psi}_2}
\newcommand\tpsii{\tilde{\psi}_i}

\newcommand\Ms{M_*}
\newcommand\Mbh{M_{\rm {BH}}}
\newcommand\Mg{M_{\rm g}}

\newcommand\MR{{\cal R}}
\newcommand\MRc{{\cal R}_{\rm c}}
\newcommand\MReq{{\cal R}_{\frac{\upi}{2}}}
\newcommand\MRinf{{\cal R}_{\infty}}
\newcommand\MRint{{\cal R}_{\rm int}}
\newcommand\Rm{{\cal R}_{\rm m}}

\newcommand\sigsb{\sigma_{\rm BH}}
\newcommand\sigsg{\sigma_{\rm g}}
\newcommand\sigs{\sigma_*}
\newcommand\sigphi{\sigma_{\varphi}}
\newcommand\vphib{\overline{v_\varphi}}
\newcommand\vphisb{\overline{\vphi^2}}

\newcommand\vmedio{\overline{\bf{v}}}
\newcommand\vlos{v_{\rm los}}
\newcommand\siglos{\sigma_{\rm los}}
\newcommand\nv{{\bf n}}
\newcommand\vv{{\bf v}}
\newcommand\vp{v_{\rm p}}
\newcommand\Vp{V_{\rm p}}
\newcommand\sigij{\sigma_{ij}}

\newcommand\Dels{\Delta_*}
\newcommand\Delsg{\Delta_{\rm g}}
\newcommand\Delsbh{\Delta_{\rm BH}}

\newcommand\Ks{K_*}
\newcommand\Ws{W_*}
\newcommand\Wss{W_{**}}
\newcommand\Wg{W_{\rm *g}}
\newcommand\Wbh{W_{\rm *BH}}
\newcommand\Wdm{W_{\rm *DM}}
\newcommand\Us{U_*}
\newcommand\Uss{U_{**}}
\newcommand\Ug{U_{\rm *g}}
\newcommand\Ubh{U_{\rm *BH}}
\newcommand\Udm{U_{\rm *DM}}
\newcommand\Bg{B_{*\rm g}}
\newcommand\Un{U_{\rm n}}

\newcommand\sigp{\sigma_{\rm p}}

\newcommand\JJ{{\rm JJ}}
\newcommand\Jt{{\rm J3}}
\newcommand\JJe{{\rm JJe}}
\newcommand\Jte{{\rm J3e}}

\newcommand\AD{{\rm AD}}
\newcommand\vphi{v_{\varphi}}
\newcommand\vz{v_z}

\newcommand\vR{v_R}
\newcommand\Hcsi{{\cal H}}

\newcommand\tvcz{\tilde{v}^2_{{\rm g}0}}
\newcommand\tvcu{\tilde{v}^2_{{\rm g}1}}
\newcommand\vphimq{\overline{v_{\varphi}}^{\hspace{0.2mm}2}}

\newcommand\lapn{\tilde{\nabla}^2}

\newcommand\ag{a_{\rm g}}
\newcommand\aBH{a_{\rm BH}}

\newcommand\Deltac{{\mathcal{D}}_{\rm c}}
\newcommand\Jc{J_{\rm c}}
\newcommand\Rin{R_{\rm in}}

\newcommand\tX{\tilde{X}}
\newcommand\tY{\tilde{Y}}

\newcommand\kvert{\kappa_z}
\newcommand\krad{\kappa_R}

\newcommand\kverttz{\tilde{\kappa}_{z0}}
\newcommand\kverttu{\tilde{\kappa}_{z1}}
\newcommand\kverttd{\tilde{\kappa}_{z2}}
\newcommand\kvertti{\tilde{\kappa}_{zi}}

\newcommand\kradtz{\tilde{\kappa}_{R0}}
\newcommand\kradtu{\tilde{\kappa}_{R1}}
\newcommand\kradti{\tilde{\kappa}_{Ri}}
\newcommand\kn{\kappa_{\rm n}}

\title[Two-component galaxy models with a central BH]{Two-component
  galaxy models with a central BH -- II. The ellipsoidal case}

\author[L. Ciotti, A. Mancino, S. Pellegrini \& A. Ziaee~Lorzad]{Luca Ciotti$^1$, Antonio Mancino$^{1,2}$, Silvia Pellegrini$^{1,2}$ \& Azadeh Ziaee~Lorzad$^1$ 
\\
$^1$Department of Physics and Astronomy, University of Bologna, via Gobetti 93/3, 40129 Bologna, Italy
\\
$^2$Istituto Nazionale di Astrofisica (INAF), Osservatorio di Astrofisica e Scienza dello Spazio di Bologna (OAS), Via Gobetti 93/3, Bologna 40129, Italy}

\date{Accepted 2020 October 19. Received 2020 September 29; in original form 2020 July 31}

\begin{document}
\maketitle


\begin{abstract}
  
\noindent
Recently, two-component spherical galaxy models have been presented, where the stellar profile
is described by a Jaffe law, and the total density by another Jaffe law, or by an $r^{-3}$ law 
at large radii. We extend these two families to their ellipsoidal axisymmetric counterparts: the 
JJe and J3e models. The total and stellar density distributions can have different flattenings 
and scale lengths, and the dark matter halo is defined by difference. First, the analytical 
conditions required to have a nowhere negative dark matter halo density are derived. The Jeans 
equations for the stellar component are then solved analytically, in the limit of small 
flattenings, also in presence of a central BH. The azimuthal velocity dispersion anisotropy is 
described by the Satoh $k$-decomposition. Finally, we present the analytical formulae for 
velocity fields near the center and at large radii, together with the various terms entering 
the Virial Theorem. The JJe and J3e models can be useful in a number of theoretical applications,
e.g. to explore the role of the various parameters (flattening, relative scale lengths, mass 
ratios, rotational support) in determining the behavior of the stellar kinematical fields before 
performing more time-expensive integrations with specific galaxy models, to test codes of stellar 
dynamics, and in numerical simulations of gas flows in galaxies.
\end{abstract}


  \section{Introduction}\label{sec:Intro}

  Axially symmetrical models of galaxies are useful tools in Stellar 
  Dynamics (see e.g. BT08), and are often adopted to investigate the 
  presence of dark matter halos (hereafter DM), or central black holes 
  (hereafter BHs), or to study the orbital structure of these systems. 

  In this paper we extend to the ellipsoidal axisymmetric case two 
  families of two-component (stars plus DM) spherical galaxy models 
  that have been recently presented. In the first family of spherical 
  models (JJ models; Ciotti \& Ziaee Lorzad 2018, hereafter CZ18) the 
  stellar density profile is described by a Jaffe (1983) law, while 
  the total is another spherical Jaffe model of larger total mass and 
  different scale length. In the second family (J3 models; Ciotti, 
  Mancino \& Pellegrini 2019, hereafter CMP19) the stellar density 
  follows again a Jaffe model, while the total is a spherical density 
  profile with a logarithmic slope equal to $-3$ at large radii. 
  Therefore, the total mass is finite in the JJ models, and infinite 
  in the J3 ones. In addition, as supermassive BHs with a mass of the 
  order of $\Mbh\simeq 10^{-3}\Ms$ are generally found at the center 
  of stellar spheroids of total mass $\Ms$ (see e.g.  Magorrian et al. 
  1988; Kormendy \& Ho 2013), in both models a BH is added at the 
  center of the galaxy.

  In CZ18 it was shown that it is always possible to choose a total 
  mass so that the DM halo resulting from the difference between the 
  total and the stellar density distributions reproduces remarkably 
  well the Navarro-Frenk-White profile (Navarro, Frenk \& White 1997, 
  hereafter NFW) in the inner region. This interesting possibility was 
  further improved in CMP19, where it was proved that the DM halo in 
  the so-called {\it minimum halo} model can be tuned to reproduce 
  very well the NFW profile over the {\it whole} radial range. 
  Summarizing, JJ and J3 models present several interesting features, 
  such as analytical simplicity, flexibility in the choice of the 
  structural parameters, realistic stellar and DM density profiles, 
  and fully analytical solutions for the Jeans equations even in 
  presence of a central BH. It is then natural to explore the 
  possibility of a generalization of these spherical models to 
  ellipsoidal (axisymmetrical) shapes: we call the new models JJe and 
  J3e, respectively. Some additional considerations are in order. The 
  first concerns the positivity of the DM halo obtained as the 
  difference of two ellipsoidal distributions with different 
  flattenings and scale lengths. Quite surprisingly, we find that the 
  problem can be solved analytically, and the constraints on the model 
  parameters in order to have a positive DM can be expressed via 
  extremely simple algebraic relations. As a consequence, the 
  positivity problem in JJe and J3e models does not require numerical 
  investigations. The second consideration is about the solution of 
  the Jeans equations. As shown in CZ18 and CMP19, in the spherical 
  case they can be solved analytically, but of course in ellipsoidal 
  models this cannot be expected to be true. In general, to solve 
  them for JJe and J3e models requires the use of numerical codes 
  (see Caravita et al. 2020, in preparation). However, in the limit 
  of small flattening, density and potential of ellipsoidal 
  distributions can be expanded at the desired order in the 
  flattening by using the {\it homoeoidal expansion} method. Here we 
  show that in this limit the Jeans equations for the homoeoidally 
  expanded JJe and J3e models can be solved analytically (although, 
  not unexpectedly, the formulae are more complicated than that in 
  the sperical case). This possibility to study and plot the resulting 
  kinematical fields, and also to have the quantities entering the 
  Virial Theorem expressed in analytical form for realistic ellipsoidal 
  two-component models, without the need for resorting to numerical 
  time-expensive integrations, is a very useful property of JJe and J3e 
  models. 

  The models here introduced, in addition to the standard applications, 
  can be useful in the building of hydrostatic, barotropic and baroclinic 
  models for hot rotating atmospheres in galaxies (see e.g. Barnab\`e et. 
  al 2006). Moreover, they can be adopted in hydrodynamical simulations 
  of gas flows in galaxies, where the stellar velocity fields are major 
  ingredients in the description of the energy and momentum source terms 
  due to the evolving stellar populations (see e.g. Posacki et al. 2013; 
  Negri et. al 2014).

  The paper is organized as follows. In Section \ref{sec:Models} the main 
  structural properties of the models are presented, and in Section 
  \ref{sec:JEs} we set up and discuss the associated Jeans equations. In 
  Section \ref{sec:Solution} the solution of the Jeans equations is 
  presented, while in Section \ref{sec:Proj} the asymptotic behaviours of 
  the projected velocity profile at small and large radii are discussed. 
  In Section \ref{sec:VT} the Virial Theorem is presented, and the global 
  energetics is explicitly calculated. The main results are finally 
  summarized in Section \ref{sec:Conclusions}, while the Appendices 
  contain technical details and formulae.

  \section{The models}\label{sec:Models}

  The ellipsoidal JJ models (hereafter, JJe models) and the ellipsoidal
  J3 models (hereafter, J3e models) are the natural generalization of 
  the spherically symmetric JJ and J3 models introduced and fully 
  discussed in CZ18 and CMP19, respectively. It is however useful to 
  recall how these spherical models are defined. The stellar density 
  profile 
  \begin{equation}
        \rhos(r)=\frac{\rhon}{s^2(1+s)^2}
  \label{eq:rhos}
  \end{equation}
  is the same for the two families, where
  \begin{equation}
        \rhon \equiv \frac{\Ms}{4\upi\rs^3}, 
        \qquad\quad 
        s \equiv \frac{r}{\rs},
  \label{eq:rhon_rs}
  \end{equation}
  and $\Ms$ and $\rs$ are the total stellar mass and the stellar scale 
  length, respectively. The normalization potential and the BH-to-stellar 
  mass ratio are defined as
  \begin{equation}
        \Psin \equiv \frac{G\Ms}{\rs}, 
        \qquad\quad 
        \mu \equiv \frac{\Mbh}{\Ms}.
  \end{equation} 
  Following CZ18 and CMP19 the total density profiles can be written as
  \begin{equation}
        \rhog(r)=\rhon\hspace{-0.1mm}\times
        \begin{cases}
              \hspace{0.05cm}\displaystyle{\frac{\MR\xi}{s^2(\xi+s)^2}}, \hspace{0.952cm} (\JJ),
              \\[15pt]
              \hspace{0.05cm}\displaystyle{\frac{\MR}{s^2(\xi+s)}}, \hspace{1.1cm} (\Jt),
        \end{cases}
  \label{eq:rhog_sph}
  \end{equation} 
  where $\xi \equiv \rg/\rs$ is the galaxy scale length $\rg$ in units 
  of $\rs$, and $\MR$ is a measure of the total-to-stellar density: in 
  particular, for JJ models $\MR=\Mg/\Ms$, while for J3 models $\MR/\xi$ 
  can be interpreted as the ratio of the total over the stellar density 
  at the center. Of course, this latter interpretation holds also for 
  the JJ models. The cumulative galactic mass inside a sphere of radius 
  $r$ reads
  \begin{equation}
        \Mg(r)=\Ms\hspace{-0.2mm}\times
        \begin{cases}
              \hspace{0.05cm}\displaystyle{\frac{\MR s}{\xi+s}}, \hspace{1.555cm} (\JJ),
              \\[15pt]
              \hspace{0.05cm}\displaystyle{\MR\ln\frac{\xi+s}{\xi}}, \hspace{0.9cm} (\Jt).
        \end{cases}
  \end{equation}
  Note that, while $\Mg(r)$ tends to a finite value in the case of JJ models, 
  for J3 models it diverges logarithmically. Finally, the (relative) galaxy 
  potential is given by
  \begin{equation}
        \Psig(r)=\Psin\hspace{-0.2mm}\times
        \begin{cases}
              \hspace{0.05cm}\displaystyle{\frac{\MR}{\xi}\ln\frac{\xi+s}{s}}, \hspace{2.52cm} (\JJ),
              \\[15pt]
              \hspace{0.05cm}\displaystyle{\frac{\MR}{\xi}\ln\frac{\xi+s}{s}+\frac{\MR}{s}\ln\frac{\xi+s}{\xi}}, \hspace{0.6cm} (\Jt).
        \end{cases}
  \end{equation}
  In the present approach, the DM halo density distribution is obtained 
  as $\rhoDM(r)=\rhog(r)-\rhos(r)$.

  We can now introduce the new ellipsoidal models discussed in this paper.
  The stellar component of JJe and J3e models is given by the mass-conserving 
  ellipsoidal generalization of equation \eqref{eq:rhos},
  \begin{equation}
        \rhos(R,z)=\frac{\rhon}{\qs \ms^2(1+\ms)^2}, 
        \qquad
        \ms^2=\Rtil^2+\frac{\ztil^2}{\qs^2},
  \label{eq:rhos(ms)}
  \end{equation} 
  where $\qs=1-\etas$ measures the flattening of the density distribution, 
  and $\Rtil\equiv R/\rs$ and $\ztil\equiv z/\rs$ are the dimensionless 
  cylindrical coordinates. From volume integration of equation 
  \eqref{eq:rhos(ms)} the independence of the total mass $\Ms$ on $\qs$ 
  can be immediately verified. In analogy with equation \eqref{eq:rhog_sph}, 
  for the total galaxy density profile we assume
  \begin{equation}
        \rhog(R,z)=\rhon\hspace{-0.1mm}\times
        \begin{cases}
              \hspace{0.05cm}\displaystyle{\frac{\MR\xi}{\qg\mg^2(\xi+\mg)^2}}, \hspace{1cm} (\JJe),
              \\[15pt]
              \hspace{0.05cm}\displaystyle{\frac{\MR}{\qg\mg^2(\xi+\mg)}}, \hspace{1.145cm} (\Jte),
        \end{cases}
  \label{eq:rhog(mg)}
  \end{equation}
  where $\mg^2=\Rtil^2+\ztil^2/\qg^2$, and $\qg=1-\etag$ is the axial 
  ratio of the total density profile. Of course, when $\qs=\qg=1$, JJe 
  and J3e models reduce respectively to JJ and J3 models. As expected, 
  the total mass $\Mg(R,z)$ converges to a finite $\Mg$ in JJe models, 
  and diverges in J3e case. It is important to note that for $\qg\neq\qs$, 
  i.e. in the case of different flattenings for the total and stellar 
  densities, $\rhoDM$ is {\it not} stratified on ellipsoidal surfaces.

  \begin{figure*}
        \includegraphics[width=0.492\linewidth]{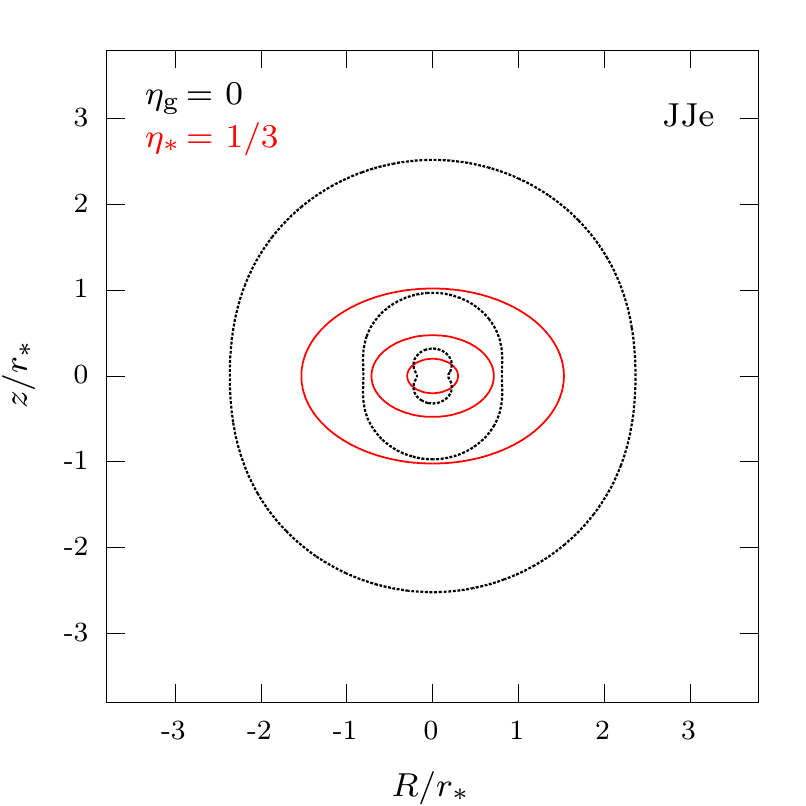}
        \hspace{0.13mm}
        \includegraphics[width=0.492\linewidth]{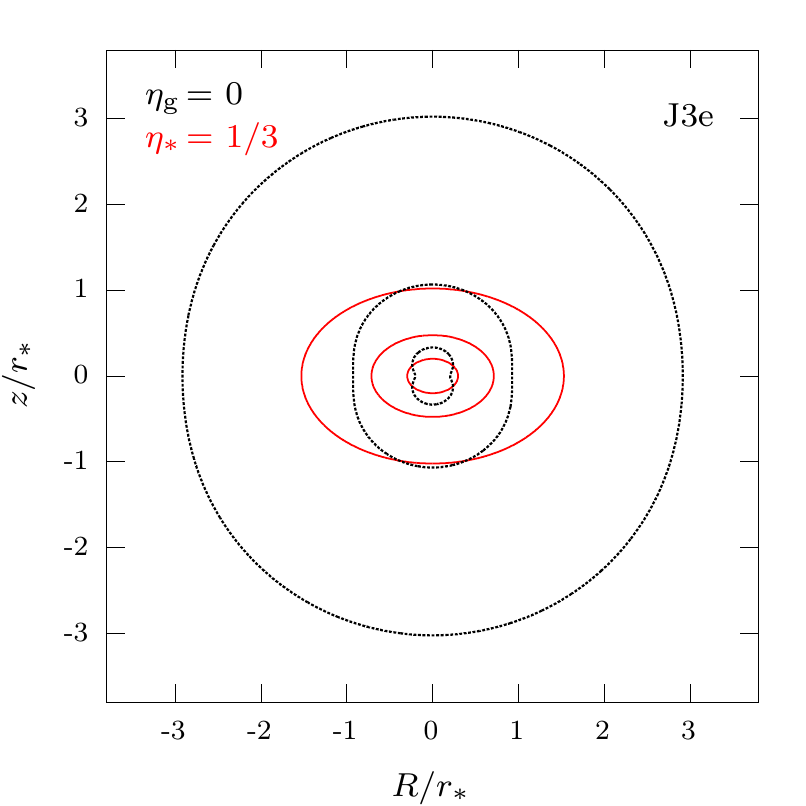}
        \\
        \hspace{0.66mm}\includegraphics[width=0.492\linewidth]{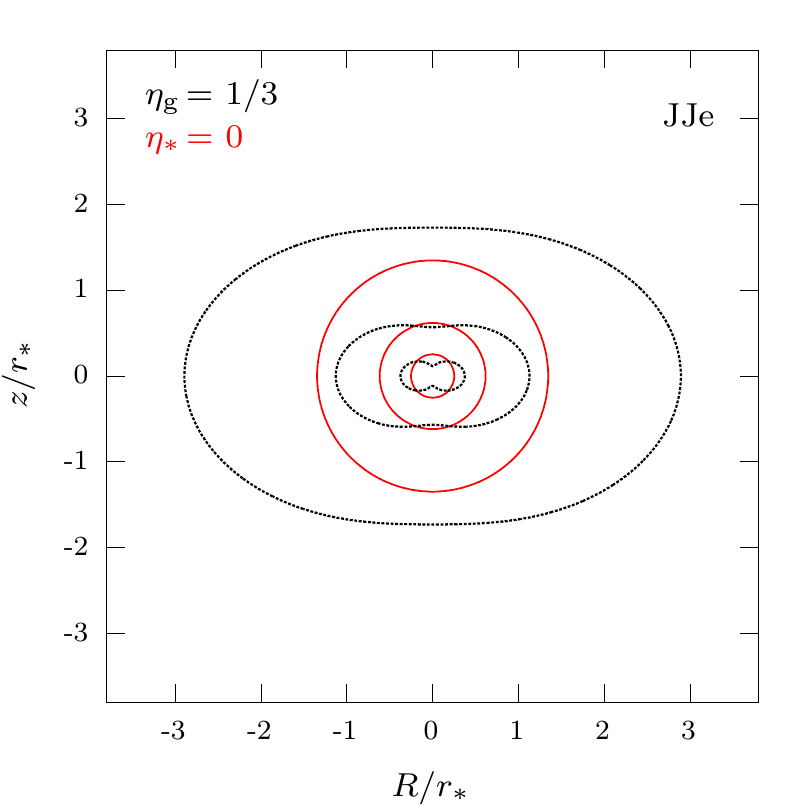}
        \hspace{0.13mm}
        \includegraphics[width=0.492\linewidth]{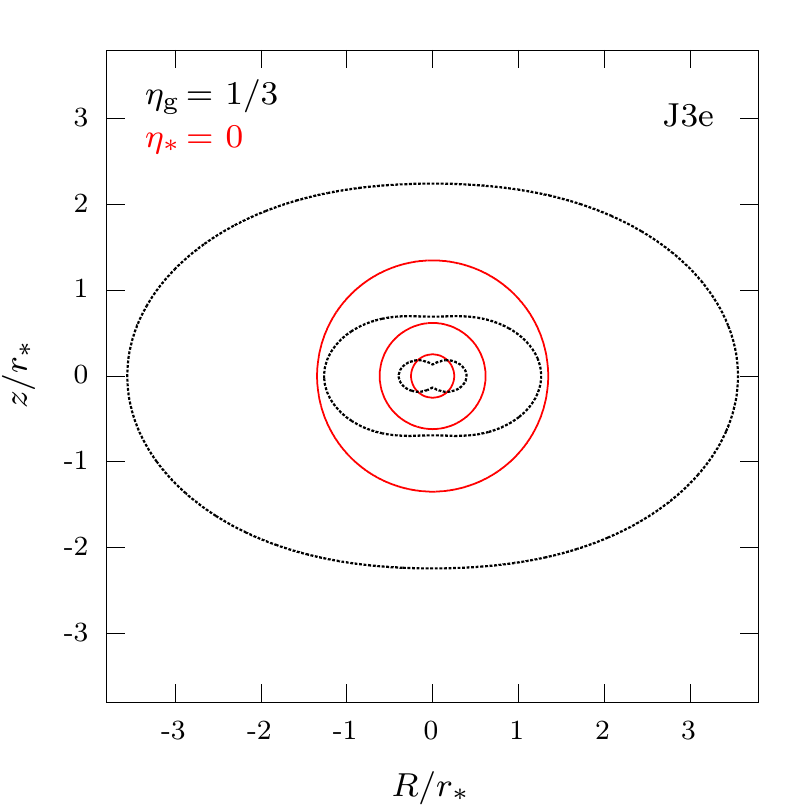}
        \caption{Isodensity contours of the stellar (solid) and DM (dotted) density distributions, 
                 for JJe (left) and J3e (right) models. The densities are normalized to $\rhon$, 
                 and the lengths to $\rs$. The contours correspond to values of $10$, $1$, and 
                 $10^{-1}$, from inside to outside. Top panels: the stellar distribution is flatter 
                 than the total, with $\etas=1/3$, while the galaxy is spherical. Bottom panels: 
                 the stellar distribution is spherical, while the total is flatter, with $\etag=1/3$. 
                 Both JJe and J3e models are minimum halo models with $\xi=5$, for which $\Rm = 15/2$ 
                 (see equations \protect\ref{eq:pos_cond_JJe} and \protect\ref{eq:pos_cond_J3e}).}
  \label{fig:iso_dens_JJe}
  \end{figure*}

  The projected stellar density associated with equation \eqref{eq:rhos(ms)} 
  can be easily obtained from ellipsoidal projection. In particular, when 
  the line-of-sight is inclined by an angle $i$ measured from the 
  $z$-axis (for example, with rotation around the $y$-axis), the 
  projected density is
  \begin{equation}
        \Sigma_*(\ell\hspace{0.3mm})=\frac{\Ms}{\rs^2\hspace{0.3mm}\qs\!\hspace{0.5mm}(i)}f(\ell\hspace{0.3mm}),
        \hspace{5mm} 
        \qs\!\hspace{0.5mm}(i)\equiv\sqrt{\cos^2\hspace{-0.3mm}i+\qs^2\sin^2\hspace{-0.3mm}i},
  \end{equation}
  (see e.g. Riciputi et al. 2005), where the function $f(\ell)$ is given 
  in equation 6 of CZ18, and the isodensity 
  ($\hspace{0.15mm}\ell\hspace{0.45mm}$) in the projection plane 
  $(X,Y)$ is
  \begin{equation}
        \ell^2=\frac{X^2}{\rs^2\hspace{0.3mm}\qs^2\!\hspace{0.5mm}(i)}+\frac{Y^2}{\rs^2}, 
  \end{equation}
  so that $\qs(i)$ is the ``isophotal'' flattening (see Section \ref{sec:Proj}). The projected 
  stellar mass contained inside the ellipse defined by $\ell$ is 
  \begin{equation}
        M_{\rm p*}(\ell\hspace{0.3mm})=\Ms\!\times g(\ell\hspace{0.3mm}),
  \end{equation}
  where $g(\ell\hspace{0.3mm})$ is again the same function as in the 
  spherical case (see CZ18, equation 9). In particular, the effective 
  ellipse corresponds to $g(\ell_e)=1/2$, i.e. $\ell_e\simeq 0.7447$, 
  with a circularized radius (i.e., the radius of the circle in the 
  projection plane with the same area of the effective ellipse) given 
  by $\langle\reff\rangle\simeq 0.7447\,\rs\sqrt{\qs(i)}$.

  \subsection{The dark matter halo: positivity} \label{sec:DMhalo_pos}

  \begin{figure*}
        \includegraphics[width=0.49\linewidth]{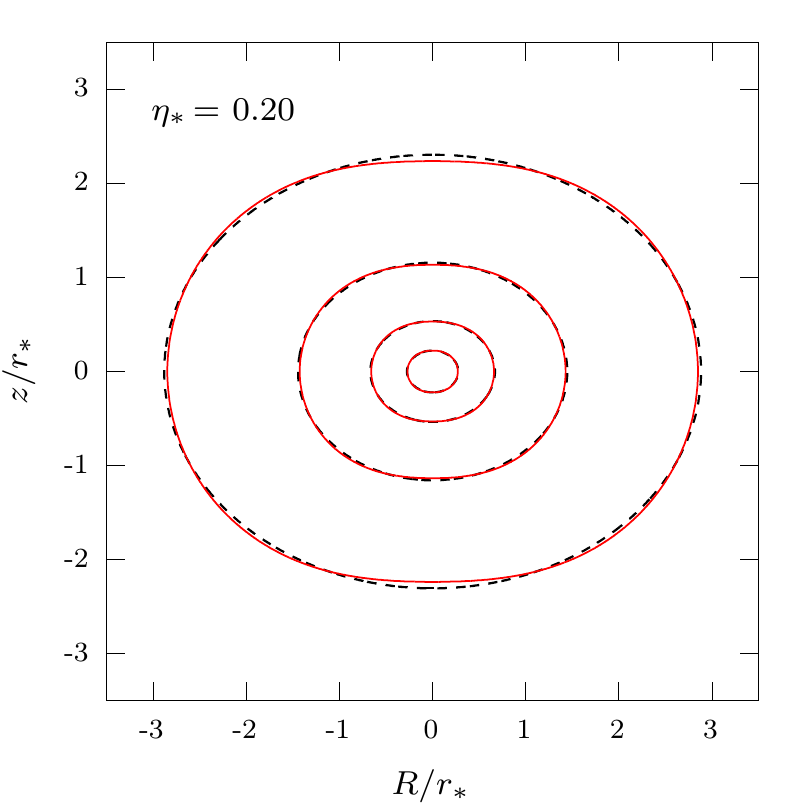}
        \hspace{0.13mm}
        \includegraphics[width=0.49\linewidth]{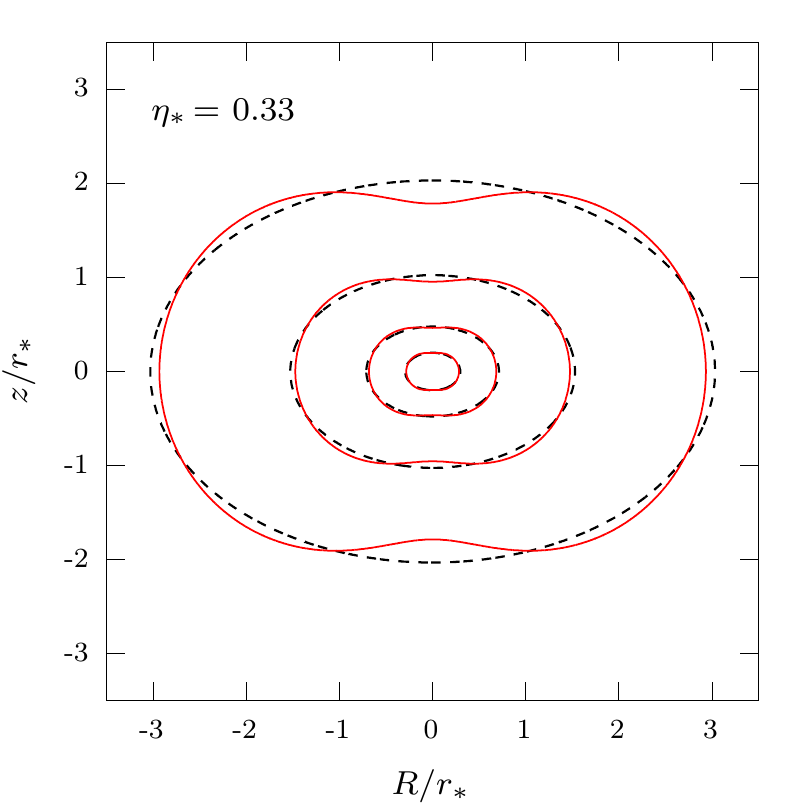}
        \vspace{0.5mm}
        \caption{Stellar isodensity contours for two values of $\etas$, for the true model 
                 (dashed), and the homoeoidal expansion (solid); the densities are normalized 
                 to $\rhon$ and the lengths to $\rs$. The contours correspond to values of 
                 $10$, $1$, $10^{-1}$, $10^{-2}$, from inside to outside.}
  \label{fig:rhos_trueVShom}
  \end{figure*}

  As the DM component is given by a difference of two density 
  distributions,a preliminary study of the positivity of its density 
  $\rhoDM$ as a function of the parameters is in order. We follow 
  the similar approach already discussed for spherical models in CZ18 
  and CMP19; of course, the situation is now more complicated, due to 
  the possible different shape of the total and stellar distributions. 
  Quite surprisingly, we find that the discussion can be carried out 
  analytically. We begin by considering the case of two-component 
  ellipsoidal $\gamma$ models (see Dehnen 1993; Tremaine et al. 
  1994), where the DM density profile is defined as
  \begin{equation}
        \rhoDM(R,z)=\frac{(3-\gamma)\rhon\MR\xi}{\qg\mg^{\gamma}(\xi+\mg)^{4-\gamma}} 
                    -\frac{(3-\gamma)\rhon}{\qs\ms^{\gamma}(1+\ms)^{4-\gamma}}.
  \label{eq:rhoDM_gammagamma}
  \end{equation}
  Then, JJe models are the $\gamma=2$ case. In Appendix \ref{app:pos} 
  we show that the positivity of $\rhoDM$ requires $\MR\geq \Rm$, with
  \begin{equation}
        \Rm(\JJe)=
        \begin{cases}
              \hspace{0.05cm}\displaystyle{\frac{\qg}{\qs} \max\!\hspace{0.2mm}\left(\frac{1}{\xi},\,\xi\right)}, \hspace{1.05cm} \qs \leq \qg,
              \\[15pt]
              \hspace{0.05cm}\displaystyle{\frac{\qs}{\qg}\max\!\hspace{0.2mm}\left(\frac{\qs^2}{\qg^2\xi},\,\xi\right)}, \hspace{0.75cm} \qs \geq \qg.
        \end{cases}
  \label{eq:pos_cond_JJe}
  \end{equation}
  Note that, once $\qs$ and $\qg$ have been chosen, 
  $\Rm(\JJe) \geq 1$ for every value of $\xi>0$. A model 
  with $\MR=\Rm$ is called a {\it minimum halo model}. 
  Clearly, when $\qg=\qs$ the positivity conditions reduce 
  to that of spherical JJ models (see CZ18, equation 18).\\ 
  The positivity condition for the DM component of J3e 
  models is instead given by
  \begin{equation}
        \Rm(\Jte)=\max\!\hspace{0.35mm}\big(\MRc,\MR_0,\MReq,\MRint\big),
  \label{eq:pos_cond_J3e}
  \end{equation}
  where $\MRc$, $\MR_0$, $\MReq$, and $\MRint$ are given in 
  Appendix \ref{app:pos}. We may notice that, at variance with 
  the previous case, $\Rm(\Jte)$ can be less than unity for 
  some choice of the parameters; this is not surprisingly, since 
  the positivity condition for J3e models {\it must} reduce to 
  that of spherical J3 models for $\qs=\qg=0$, which gives 
  $\Rm<1$ when $\xi<1$ (see equation 16 in CMP19).

  For illustrative purposes, in Fig. \ref{fig:iso_dens_JJe} we 
  show the isodensity contours of the stellar and DM density 
  profiles in the meridional plane, for four minimum halo galaxy 
  models. In the top panels the galaxy is spherical ($\etag=0$), 
  while $\etas=1/3$; in the bottom panels the galaxy is flatter 
  ($\etag=1/3$), while the stellar density is spherical. Note 
  that, as expected, for $\etag =0$ the DM distribution is 
  elongated along the $z$-axis, with a prolate-like shape. From 
  the results in Appendix \ref{app:pos}, one has that a negative 
  DM density is obtained for a total mass below the minimum halo 
  mass. Finally, note how, at any fixed distance from the galactic 
  center (but especially outside $\rs$), the DM density is larger 
  for J3e than for JJe models.

  \subsection{Homoeoidal expansion} \label{sec:Hom_Exp}

  As pointed out in the Introduction, one of the main ideas behind 
  this work is to combine the approach of model construction based 
  on the assignment of the total and stellar profiles (as carried 
  out for spherical models in CMZ09, CZ18, and CMP19) with the 
  homoeoidal expansion technique (see CB05), a methods that allows 
  to describe, in a tractable way, ellipsoidal models in the limit 
  of {\it small} flattening. Of course, the models here presented 
  can also be investigated in the case of {\it finite} flattenings 
  by using a numerical approach (see Caravita et al. 2020, in 
  preparation), and the comparison of analytical and numerical 
  results is a useful sanity check for both methods. Finally, the 
  present approach, combining the merits of model difference and 
  analytical tractability, is not completely new; in particular, 
  we recall the seminal paper by Evans (1993).

  Before presenting the analytical solution of the Jeans equations 
  for the JJe and J3e models (see Sections \ref{sec:JEs} and 
  \ref{sec:Solution}), we now consider the homoeoidal expansion at
  fixed mass (the so-called {\it constrained} expansion), as a 
  function of the two flattenings $\etas$ and $\etag$. The formulae 
  are obtained from Appendix \ref{app:Hom_Exp}, where we also show that, 
  in order to have physically acceptable stellar density, $\etas \leq 1/3$; 
  when considering the total density, instead, equation 
  \eqref{eq:pos_cond_general} shows that $\etag \leq 1/3$ for JJe 
  models, while $\etag \leq 1/2$ for J3e models. 

  \begin{figure*}
        \includegraphics[width=0.493\linewidth]{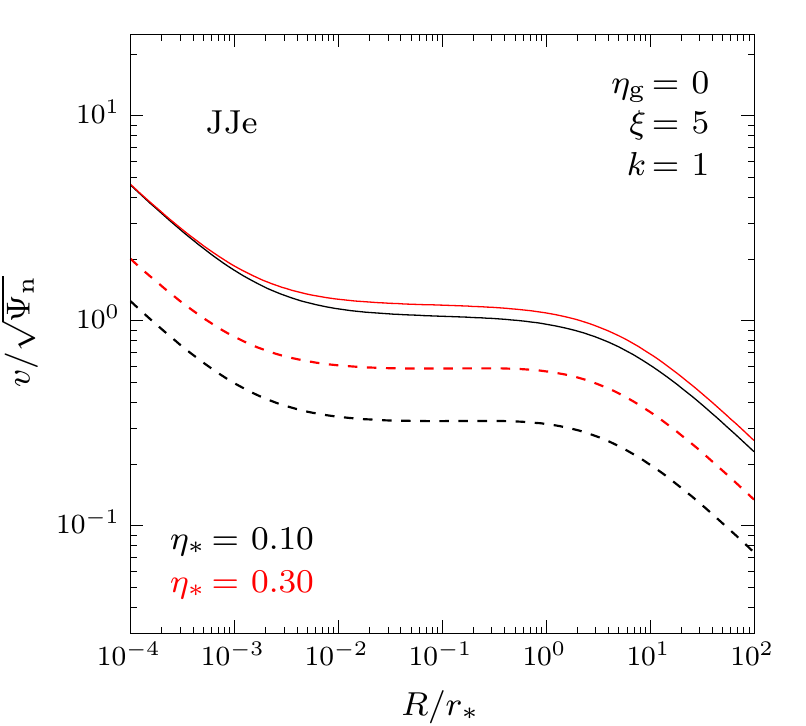}
        \hspace{0.8mm}
        \includegraphics[width=0.493\linewidth]{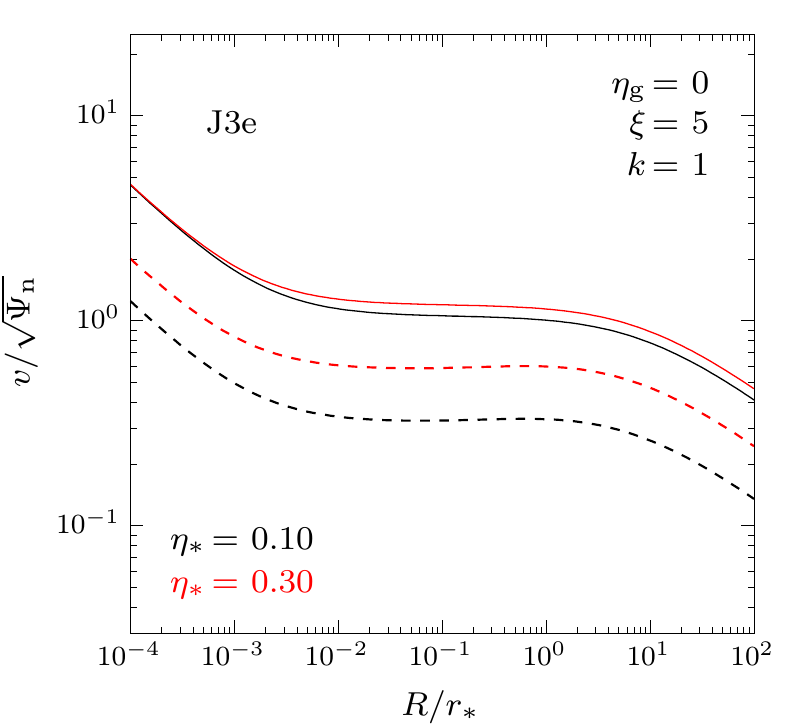}
        \vspace{-3mm}
        \caption{Radial trends of $\vcirc(R\hspace{0.25mm})$ (solid lines) and 
                 $\vphib\hspace{0.4mm}(R,0)$ (dashed lines), normalized to $\sqrt{\Psin}$. 
                 Left: JJe models; right: J3e models. Both panels refer to a spherical 
                 galaxy, with a central BH with $\mu=0.002$, isotropic orbits (i.e., $k=1$), 
                 and a minimum halo with $\xi=5$; from equations \protect\eqref{eq:pos_cond_JJe} and
                 \protect\eqref{eq:pos_cond_J3e}, $\Rm=50/9$ when $\etas=0.10$, and $\Rm=50/7$ when 
                 $\etas=0.30$.}
  \label{fig:vcVSvphimRz}
  \end{figure*}

  The expansion of the total density up to linear terms in the 
  flattenings $\etag \to 0$ reads
  \begin{equation}
        \rhog(R,z)=\rhon\MR\hspace{0.25mm} 
        \Big[
        \hspace{0.2mm}\rtgz(s)+\etag\rtgu(s)+\etag\Rtil^2\rtgd(s)
        \Big],
  \label{eq:rhog_exp}
  \end{equation}
  where $s=r/\rs$, $\Rtil = R/\rs$, and the 
  general expression of the three 
  spherical functions is given in Appendix \ref{app:Hom_Exp}. 
  As will be discussed in Section \ref{sec:JEs}, there are two 
  different interpretations of the expansion above: one as the 
  linearized expansion of a true ellipsoidal model, and the other 
  as a genuine density distribution with arbitrary values of 
  $\etag$, provided that positivity is assured. We shall return 
  to this point. 

  For JJe and J3e models, the three dimensionless functions are 
  given by
  \begin{equation}
        \rtgi(\JJe)=
        \begin{cases} 
              \hspace{0.05cm}\displaystyle{\frac{\xi}{s^2(\xi+s)^2}}, & \hspace{0.3cm}  
              \\[10pt]
              \hspace{0.05cm}\displaystyle{-\hspace{0.35mm}\frac{\xi (\xi +3s)}{s^2(\xi+s)^3}}, 
              \\[10pt]
              \hspace{0.05cm}\displaystyle{\frac{2\xi(\xi+2s)}{s^4(\xi+s)^3}}, 
        \end{cases}\!
        \rtgi(\Jte)=
        \begin{cases} 
              \hspace{0.05cm}\displaystyle{\frac{1}{s^2(\xi+s)}}, & \hspace{0.3cm}  
              \\[10pt]
              \hspace{0.05cm}\displaystyle{-\hspace{0.35mm}\frac{\xi+2s}{s^2(\xi+s)^2}}, 
              \\[10pt]
              \hspace{0.05cm}\displaystyle{\frac{2\xi+3s}{s^4(\xi+s)^2}}, 
        \end{cases}
  \label{eq:rtgi_JJe_J3e}
  \end{equation}
  for $i=0,1,2$, from top to bottom, respectively. Note that 
  $\rtgu$ is everywhere negative, whereas $\rtgz$ and $\rtgd$ 
  are positive functions of $s$. 
  Finally, since the stellar 
  distribution is the same for the JJe and J3e models, and it 
  is of the same family of $\rhog$ of JJe models, its homoeoidal 
  expansion is
  \begin{equation}
        \rhos(\JJe,\Jte)=\rhon\hspace{0.25mm} 
        \Big[
        \hspace{0.2mm}\rtsz(s)+\etas\rtsu(s)+\etas\Rtil^2\rtsd(s)
        \Big],
  \label{eq:rhos_homexp}
  \end{equation}
  where the functions $\rtsi$ are obtained by setting 
  $\xi=1$ in $\rtgi(\JJe)$.
  Fig. \ref{fig:rhos_trueVShom} shows the isodensity 
  contours of the stellar density profile for two different value 
  of $\etas$. Black dashed lines correspond to the ellipsoidal 
  Jaffe model, described by equation \eqref{eq:rhos(ms)}, while 
  red solid lines refer to the homoeoidally expanded Jaffe model, 
  provided by equation \eqref{eq:rhog_exp}: note how the outermost 
  expanded contours differ from the elliptical shape as $\etas$ 
  approaches the value $1/3$.

  The potential associated with equation \eqref{eq:rhog_exp}, 
  following Appendix \ref{app:Hom_Exp}, can be written as 
  \begin{equation} 
        \Psig(R,z)=\Psin\MR\hspace{0.25mm}
        \Big[
        \tPsigz(s)+\etag\tPsigu(s)+\etag\Rtil^2\tPsigd(s)
        \Big],
  \label{eq:Psig_exp}
  \end{equation}
  where, as for the density, we recast equation 
  \eqref{eq:Psitilde_expansion} in terms of $\Rtil$. The general 
  integral expressions of the three spherical functions in the 
  equation above are given in Appendix \ref{app:Hom_Exp}; for both 
  JJe and J3e models, all these integrals are elementary, with the 
  normalized densities $\rhotil(m)=\xi/[m^2(\xi+m)^2]$ and 
  $\rhotil(m)=1/[m^2(\xi+m)]$, respectively. The final result is 
  \begin{equation}
        \tPsigi(\JJe)=
        \begin{cases} 
              \hspace{0.05cm}\displaystyle{\frac{1}{\xi}\ln\frac{\xi+ s}{s}},  
              \\[10pt]
              \hspace{0.05cm}\displaystyle{-\,\frac{s^2+2\xi s+4\xi^2}{3s^2(\xi +s)}+\frac{1}{3\xi}\ln\frac{\xi+s}{s}+\frac{4\xi^2}{3s^3}\ln\frac{\xi+s}{\xi}}, 
              \\[10pt]
              \hspace{0.05cm}\displaystyle{\frac{\xi(s+2\xi)}{s^4(\xi+s)}-\frac{2\xi^2}{s^5}\ln\frac{\xi+s}{\xi}}, 
        \end{cases}
  \label{eq:tPsigi_JJe}
  \end{equation}
  and
  \begin{equation}
        \tPsigi(\Jte)=
        \begin{cases} 
              \hspace{0.05cm}\displaystyle{\frac{1}{\xi}\ln\frac{\xi+s}{s}+\frac{1}{s}\ln\frac{\xi+s}{\xi}},  
              \\[10pt]
              \hspace{0.05cm}\displaystyle{-\,\frac{s-2\xi}{3s^2}+\frac{1}{3\xi}\ln\frac{\xi+s}{s}-\frac{2\xi^2}{3s^3}\ln\frac{\xi+s}{\xi}}, 
              \\[10pt]
              \hspace{0.05cm}\displaystyle{\frac{s-2\xi}{2s^4}+\frac{\xi^2}{s^5}\ln\frac{\xi+s}{\xi}}, 
        \end{cases}
  \label{eq:tPsigi_J3e}
  \end{equation}
  for $i=0,1,2$, from top to bottom, respectively. As a check, the 
  formulae \eqref{eq:rtgi_JJe_J3e}, \eqref{eq:tPsigi_JJe} and 
  \eqref{eq:tPsigi_J3e} have been verified to satisfy the Poisson 
  equation for the dimensionless potential-density pair 
  $(\tPsig,\trhog)$ from the linearization of equation \eqref{eq:Poisson}. 
  Finally, the total potential $\PsiT$ is obtained by adding to 
  $\Psig$ the contribution of a central BH of mass $\Mbh$, so that 
  the total potential is given by 
  \begin{equation}
        \PsiT(R,z)=\Psig(R,z) + \frac{\mu\Psin}{s}.
  \label{eq:Psi_totale}
  \end{equation}

  A useful quantity to characterize the total potential is the circular 
  velocity in the equatorial plane, given by 
  \begin{equation}
        \vcirc^2(R\hspace{0.25mm})
        =-\hspace{0.2mm}R\hspace{-0.3mm}\left(\frac{\partial\PsiT}{\partial R}\hspace{-0.1mm}\right)_{\hspace{-1mm}z=0}
        \hspace{-0.25mm}
        =\,v^2_{\rm g}(R\hspace{0.25mm})+v^2_{\rm BH}(R\hspace{0.25mm}).
  \label{eq:vcirc}
  \end{equation}
  For the homoeoidally expanded models, 
  \begin{equation}
        \vcirc^2(R\hspace{0.25mm}) 
        =\Psin\MR\hspace{0.25mm}
        \Big[
        \hspace{0.2mm}\tvcz(\Rtil\hspace{0.25mm})+\hspace{0.1mm}\etag\tvcu(\Rtil\hspace{0.25mm})
        \Big]\!
        +\frac{\mu\Psin}{\Rtil},
  \label{eq:circ_vel}
  \end{equation}
  where for JJe and J3e models the normalized functions are given by
  \begin{equation}
        \tilde{v}^2_{{\rm g}i}(\JJe)=
        \begin{cases} 
              \hspace{0.05cm}\displaystyle{\frac{1}{\xi+\Rtil}},  
              \\[12pt]
              \hspace{0.05cm}\displaystyle{\frac{\xi (\Rtil+2\xi)}{\Rtil^2(\xi +\Rtil)}-\frac{2\xi^2}{\Rtil^3}\ln\frac{\xi+\Rtil}{\xi}}, 
        \end{cases}
  \label{eq:v2gi_JJe}
  \end{equation}
  and
  \begin{equation}
        \tilde{v}^2_{{\rm g}i}(\Jte)=
        \begin{cases} 
              \hspace{0.05cm}\displaystyle{\frac{1}{\Rtil}\ln\frac{\xi+\Rtil}{\xi}},  
              \\[12pt]
              \hspace{0.05cm}\displaystyle{\frac{\Rtil-2\xi}{2\Rtil^2}+\frac{\xi^2}{\Rtil^3}\ln\frac{\xi+\Rtil}{\xi}}, 
        \end{cases}
  \label{eq:v2gi_J3e}
  \end{equation}
  for $i=0,1$, from top to bottom, respectively. As expected, 
  galaxy flattening increases the value of $\vcirc$, because the 
  gravitational field in the equatorial plane, at fixed total 
  mass and major-axis scale-length, becomes stronger for flattened 
  systems. The trend of $\vcirc(R\hspace{0.25mm})$ is shown in 
  Fig. \ref{fig:vcVSvphimRz} (solid lines) for two minimum halo 
  models with $\xi=5$, $\etag=0$, and with $\mu=0.002$ 
  (see Kormendy \& Ho 2013 for this choice of $\mu$); black lines 
  show the quite flat case with $\etas=0.10$, red lines refer to 
  the $\etas=0.30$ case. Note how, for fixed value of $R$ in the 
  external regions, $\vcirc(R\hspace{0.25mm})$ increases for 
  increasing $\etas$, due to the dependence of $\Rm$ on $\etas$; 
  near the center, instead, the BH contribution is always dominant 
  over that of the galaxy, and so 
  $\vcirc(R\hspace{0.25mm}) \propto R^{\hspace{0.3mm}-1/2}$ 
  independently on $\etas$. Finally, we note that the expansion 
  method allows for simple expressions for the {\it radial 
  epicyclic frequency} and {\it vertical epicyclic frequency}
  (see Appendix \ref{app:Hom_Exp}).

  \subsubsection{Asymptotic behaviour}

  For future use we also report the leading term of the density, 
  potential, and circular velocity in the central region and at 
  large radii, as obtained by expansion of the corresponding 
  quantities with $\Rtil=s\sin\theta$. At small radii the asymptotic 
  behaviour of density, potential, and circular velocity for JJe 
  models coincides with that of J3e models; we find 
  \begin{equation}
        \frac{\rhog}{\rhon} \sim \MR\,\frac{1+\etag(1-2\cos^2\!\theta)}{\xi s^2},
  \label{eq:rhog_exp_zero}
  \end{equation}
  \begin{equation}
        \frac{\Psig}{\Psin} \sim -\,\MR\,\frac{3+\etag}{3\xi}\ln s, 
        \qquad\qquad
        \frac{v_{\rm g}^2}{\Psin} \sim \MR\,\frac{3+\etag}{3\xi}.
  \label{eq:Psig_vg_zero}
  \end{equation} 
  As expected, in absence of a central BH, $\vcirc$ reduces to a 
  constant value depending on the models parameters. Instead, in 
  the external regions,
  \begin{equation}
        \frac{\rhog}{\rhon}\sim\MR\times
        \begin{cases}
              \hspace{0.05cm}\displaystyle{\xi\,\frac{1+\etag(1-4\cos^2\!\theta)}{s^4}}, \hspace{0.78cm} (\JJe),
              \\[12pt]
              \hspace{0.05cm}\displaystyle{\frac{1+\etag(1-3\cos^2\!\theta)}{s^3}}, \hspace{1cm} (\Jte),
        \end{cases}
  \label{eq:rhog_exp_infinity}
  \end{equation}
  \begin{equation}
        \frac{\Psig}{\Psin} \sim \frac{\MR}{s}\times
        \begin{cases}
              \hspace{0.05cm}\displaystyle{1}, \hspace{0.515cm} (\JJe),
              \\[8pt]
              \hspace{0.05cm}\displaystyle{\ln s}, \hspace{0.2cm} (\Jte),
        \end{cases}
        \quad\hspace{1mm}
        \frac{v_{\rm g}^2}{\Psin} \sim \frac{\MR}{\Rtil}\times
        \begin{cases}
              \hspace{0.05cm}\displaystyle{1}, \hspace{0.615cm} (\JJe),
              \\[8pt]
              \hspace{0.05cm}\displaystyle{\ln\Rtil}, \hspace{0.205cm} (\Jte).
        \end{cases}
  \label{eq:Psig_vc_infinity}
  \end{equation}
  Of course, the analogous expressions for the stellar density 
  are obtained by setting $\MR=\xi=1$ in equation 
  \eqref{eq:rhog_exp_zero} and in the JJe case of equation 
  \eqref{eq:rhog_exp_infinity}. Note that at variance with the 
  density, the galaxy potential $\Psig$ at large radii is spherical, 
  also for the J3e models with their divergent total mass.

  \section{The Jeans equations}\label{sec:JEs}

  For an axisymmetric density $\rhos (R,z)$ supported by a two-integrals 
  phase-space distribution function $f(E,J_z)$, the Jeans equations for 
  the stellar component are 
  \begin{equation}
        \frac{\partial\rhos\sigs^2}{\partial z}=\rhos\frac{\partial\PsiT}{\partial z},
  \label{eq:Jeans_vert}
  \end{equation}
  \begin{equation}
        \frac{\partial\rhos\sigs^2}{\partial R}-\frac{\rhos\Dels}{R}=\rhos\frac{\partial\PsiT}{\partial R},
        \qquad
        \Dels \equiv \overline{\vphi^2}-\sigs^2,
  \label{eq:Jeans_rad}
  \end{equation}
  (see e.g. BT08); $\vphi$ indicates the azimuthal component of the
  velocity ${\bf v}=(\vR,\vphi,\vz)$, and the ``bar-operator'' 
  indicates the average value over the velocity-space. These 
  equations are simplified with respect to the general case because 
  for a two-integrals system: (1) the velocity dispersion tensor is 
  aligned with the coordinate system, i.e. the phase-space average 
  of the mixed products of the velocity components vanishes,
  $\overline{\vR\vz}=\overline{\vR\vphi}=\overline{\vphi\vz}=0$; 
  (2) the only possible non-zero streaming motion is in the azimuthal 
  direction; (3) the radial and vertical velocity dispersions are equal,
  i.e. $\sigma_R^2=\overline{v_R^2}=\overline{v_z^2}=\sigma_z^2$.
  We define $\sigma_R=\sigma_z\equiv\sigs$.

  In order to split the azimuthal velocity field in its ordered 
  ($\vphib$) and random ($\sigphi$) components, we adopt the Satoh 
  (1980) $k$-decomposition
  \begin{equation} 
        \vphib=k\hspace{0.2mm}\sqrt{\Dels}\hspace{0.2mm},
  \label{eq:vphimedio_Satoh}
  \end{equation}
  so that
  \begin{equation}
        \sigphi^2 \equiv \vphisb-\vphib^{\hspace{0.2mm}2}=\sigs^2+(1-k^2)\Dels,
  \label{eq:sigmaphi_Satoh}
  \end{equation}
  where $k^2\!\leq 1$. This implicitly assumes that the phase-space 
  distribution function depends on $k$, i.e., $f=f(E,J_z;k)$. The 
  case $k=1$ corresponds to the isotropic rotator, while for $k=0$ 
  no net rotation is present, and all the flattening of $\rhos$ is 
  due to the azimuthal velocity dispersion $\sigphi$. Note that, 
  while in the Satoh decomposition $k$ is independent of position,
  in principle $k$ can be a function of $(R,z)$, bounded above by 
  the function $k_{\rm max}(R,z)$, defined by the condition 
  $\sigphi=0$ (see CP96). The $k(R,z)$ formulation can also be used 
  to add counterrotation in a controlled way (see e.g. Negri et al. 
  2014; see also Caravita et al. 2020).  

  \subsection{The vertical Jeans equation}\label{subsec:vert}

  The vertical Jeans equation \eqref{eq:Jeans_vert} is integrated at 
  fixed $R$ with the natural boundary condition of a vanishing 
  ``pressure'' for $z \to \infty$, so that
  \begin{equation}
        \rhos\sigs^2=-\!\hspace{0.2mm}\int_z^\infty \!\rhos \frac{\partial\PsiT}{\partial z'}\,dz'.
  \label{eq:Jeans_vert_sol}
  \end{equation}
  Notice that the integration variable at fixed $R$ can be 
  changed from $z'$ to $r'$, so that by adopting the expansion
  \eqref{eq:Psig_exp} the integration acts on spherical coefficients.
  Due to the relevance of equation 
  \eqref{eq:Jeans_vert_sol}, a few comments are in order before 
  proceeding to the solution.

  The first is that the integral in equation \eqref{eq:Jeans_vert_sol} 
  is given by the sum of two contributions: the effect of the galactic 
  potential $\Psig$ on the stellar component, and the effect of the 
  central BH. As the BH contribution can be calculated explicitly in 
  terms of elementary functions even considering $\rhos$ in the fully
  ellipsoidal case, equation \eqref{eq:Jeans_vert_sol} can be solved 
  with the homoeoidal approximation. 

  The second comment concerns the general case of a non-spherical 
  $\Psig$. As the integral in equation \eqref{eq:Jeans_vert_sol} is 
  performed at fixed $R$, it is natural to expand the potential in terms 
  of $\etag$, and use the ``explicit\hspace{0.2mm}-$R$ formulation'' in 
  equation \eqref{eq:Psig_exp}. The three components $\tPsigi$ are 
  spherically symmetric, so that equation \eqref{eq:Jeans_vert_sol} for 
  the fully ellipsoidal stellar density could be again expressed as three 
  integrals over the spherical radius. However, in order to obtain 
  manageable elementary expressions, we also make use of equation 
  \eqref{eq:rhos_homexp}.

  This leads to a third and final consideration. Due to the linearity of 
  Poisson's equation, the expanded potential-density pairs can be 
  interpreted in two {\it different} ways: as a genuinely non spherical 
  system of finite flattening, or as the first order expansion of the 
  ellipsoidal parent galaxy in the limit of vanishing flattening. In the 
  first case, when integrating the Jeans equation, {\it all} the terms 
  in the product under the integral should be retained, up to the 
  {\it quadratic} order in the flattenings. In the second case, only 
  {\it linear} terms in the flattenings are retained. For simplicity, 
  here we limit ourselves to the discussion of the linearized case, and 
  so we consider only the linear terms in $\etas$ and $\etag$. In turn, 
  this choice implies that, in the resulting formulae, only $R^2$ terms 
  appear explicitly. Of course, the consideration of quadratic terms in 
  the flattenings does not present special difficulties, only a larger 
  number of computations.

  \subsection{The radial Jeans equation}\label{sec:radial}

  In principle, once the vertical Jeans equation is solved, no further 
  integration is required because the quantity $\Dels$ can be obtained 
  from equation \eqref{eq:Jeans_rad} by differentiation. However, this 
  straightforward approach may produce formulae that contain non-trivial
  simplifications, and hide important properties of the solution. These 
  problems are avoided by a very elegant commutator-like formula for the 
  quantity $\Dels$. For untruncated distributions with vanishing 
  ``pressure'' at infinity, equation \eqref{eq:Jeans_rad} can be recast 
  as a commutator-like integral:
  \begin{equation}
        \frac{\rhos\Dels}{R}=\int_z^{\infty}\!
                             \left(
                             \frac{\partial\PsiT}{\partial R}\frac{\partial\rhos}{\partial z'}-
                             \frac{\partial\PsiT}{\partial z'}\frac{\partial\rhos}{\partial R}
                             \right)\!dz'
                            \equiv \hspace{0.1mm}[\hspace{0.2mm}\PsiT,\rhos].
  \label{eq:commutator}
  \end{equation}
  It is not surprising that this relation appears both in Fluid Dynamics 
  (see e.g. Rosseland 1926; Waxman 1978; Barnab\`e et al. 2006, and 
  references therein), and in Stellar Dynamics (see Hunter 1977), due to 
  the strict relation of the isotropic Jeans equations and hydrodynamic 
  equations. By looking at equation \eqref{eq:commutator}, a few important 
  considerations follow. First, for any pair of purely radial functions, 
  the commutator vanishes. Thus, for fully spherical models one has 
  $\Dels=0$, so that, in the Satoh decomposition, spherical models cannot 
  rotate, and are necessarily isotropic, independently of the value of $k$. 
  In a spherical stellar density, the only non-zero contribution to equation 
  \eqref{eq:Jeans_vert_sol} (and so to rotation in the Satoh decomposition) 
  can be produced by a non-spherical dark matter halo, and similarly, in 
  presence of a spherical total potential, rotation can arise only for 
  non-spherical stellar distributions. This is the case of the BH 
  contribution, where the only rotation is due to the departure of the stellar 
  distribution from spherical symmetry.

  Moreover, for generic spherically simmetric functions $u$ and $v$, and 
  generic function $f$ of the cylindrical radius, it holds that
  \begin{equation}
        [f(R)u(r),v(r)]=\frac{d\hspace{0.01mm}f}{dR}\int_r^{\infty}\!u(r')\hspace{0.15mm}\frac{dv(r')}{dr'}\hspace{0.25mm}dr',
  \end{equation}
  as it can be easily proved by direct computation, and with a final change 
  of integration variable from $z'$ to $r'$. Therefore, $\Dels$ in equation 
  \eqref{eq:Jeans_rad}, at linear order in the flattenings, can be produced 
  {\it only} by the effect of the $\Psi_2$ term on $\rho_0$, and by the 
  $\Psi_0$ term on $\rho_2$. The resulting $\Dels$ is proportional to $R^2$, 
  so in the Satoh decomposition $\Dels$ vanishes on the $z$-axis, for 
  sufficiently regular density distributions.

  We conclude this Section by noting that the possibility of using the Satoh 
  decomposition depends on the positivity of $\Dels$, a condition that can be 
  violated for arbitrary choices of the density components. This problem is 
  analogous to the issue encountered in the construction of rotating baroclinic 
  configurations of assigned density distribution in Fluid Dynamics; here, of 
  course, equations \eqref{eq:Jeans_vert} and \eqref{eq:Jeans_rad} are restricted 
  to the isotropic case, and the velocity dispersion is substituted by the 
  thermodynamic temperature (see Barnab\`e et al. 2006).

  \begin{figure*}
  \includegraphics[width=0.48\linewidth]{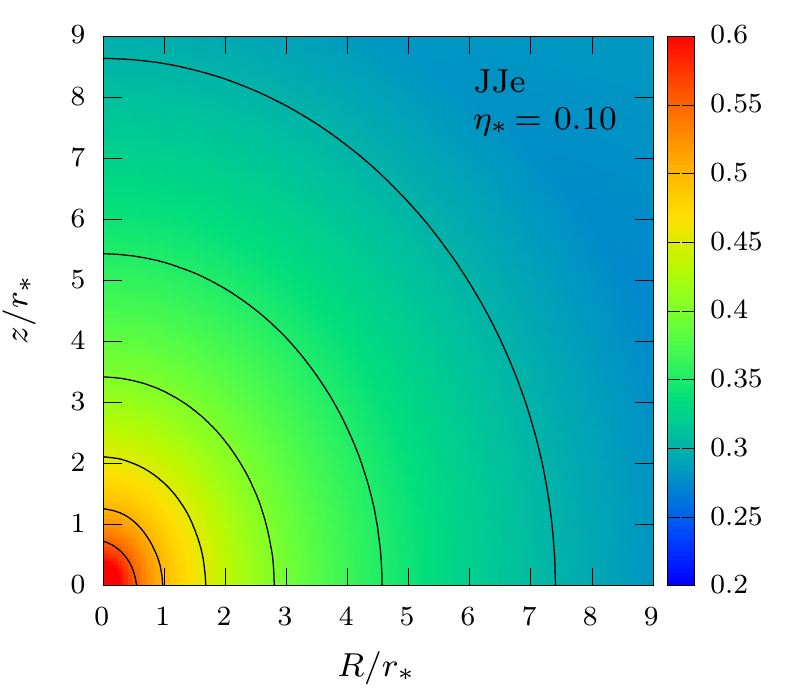}
  \hspace{2mm}
  \includegraphics[width=0.48\linewidth]{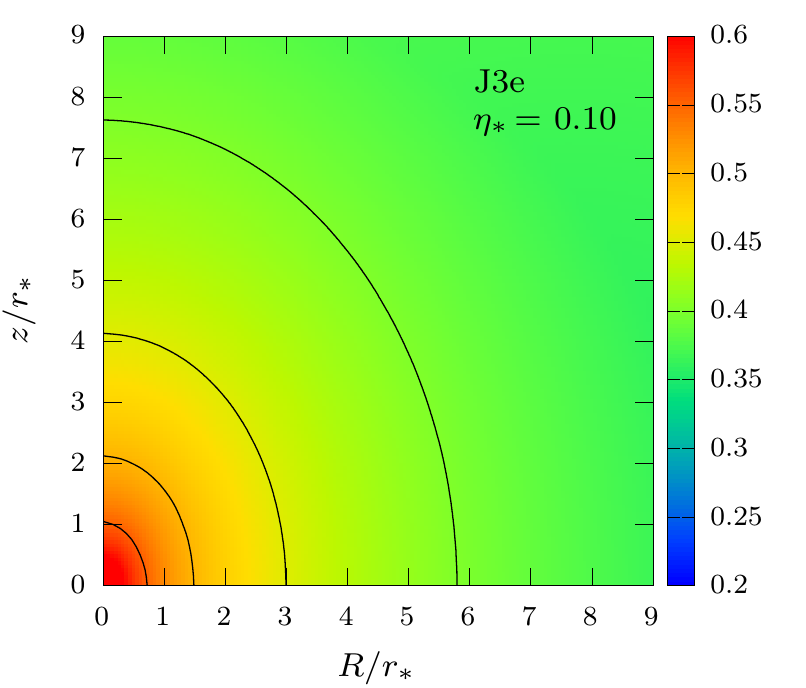}\vspace{2mm}
  \\
  \includegraphics[width=0.48\linewidth]{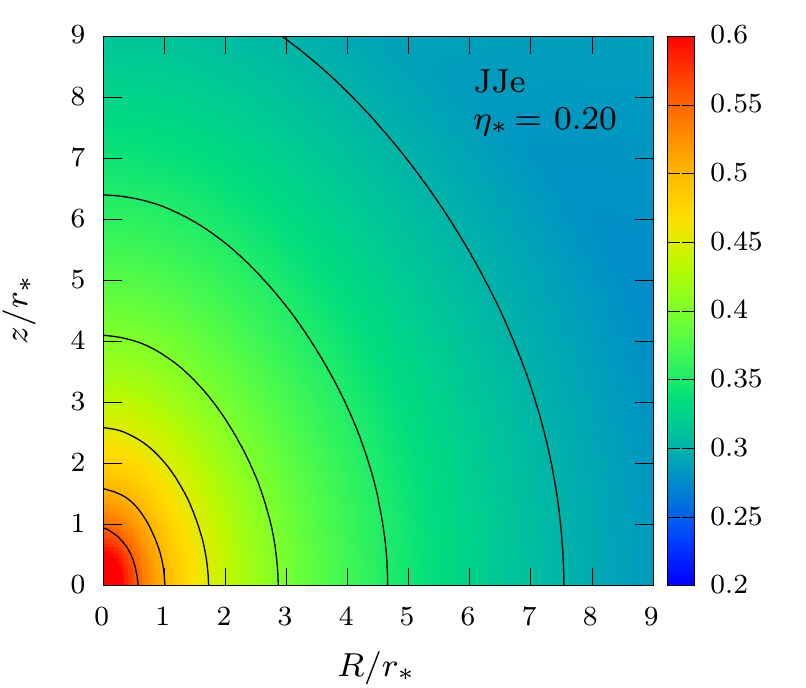}
  \hspace{2mm}
  \includegraphics[width=0.48\linewidth]{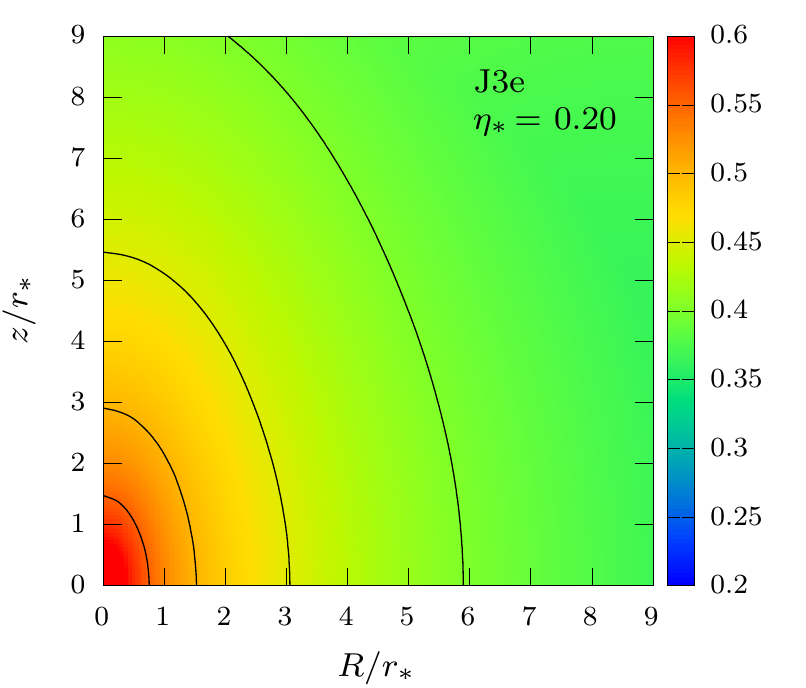}
  \vspace{0.5mm}
  \caption{Maps of $\sigs(R,z)$, normalized to $\sqrt{\Psin}$, for two representative 
           minimum halo JJe (left panels) and J3e (right panels) models with $\xi=5$, 
           for a spherical galaxy ($\etag=0$) and two values of $\etas$. 
           A central BH with $\mu=0.002$ is also present. The contour lines correspond 
           to values that, starting from $0.55$, decrease with step $0.05$ from inside 
           to outside. From equations (\protect\ref{eq:pos_cond_JJe}) and (\protect\ref{eq:pos_cond_J3e}),
           $\Rm=50/9$ when $\etas=0.10$, and $\Rm=25/4$ when $\etas=0.20$,
           for both JJe and J3e models.}
  \label{fig:sigmas}
  \end{figure*}

  \begin{figure*}
  \includegraphics[width=0.48\linewidth]{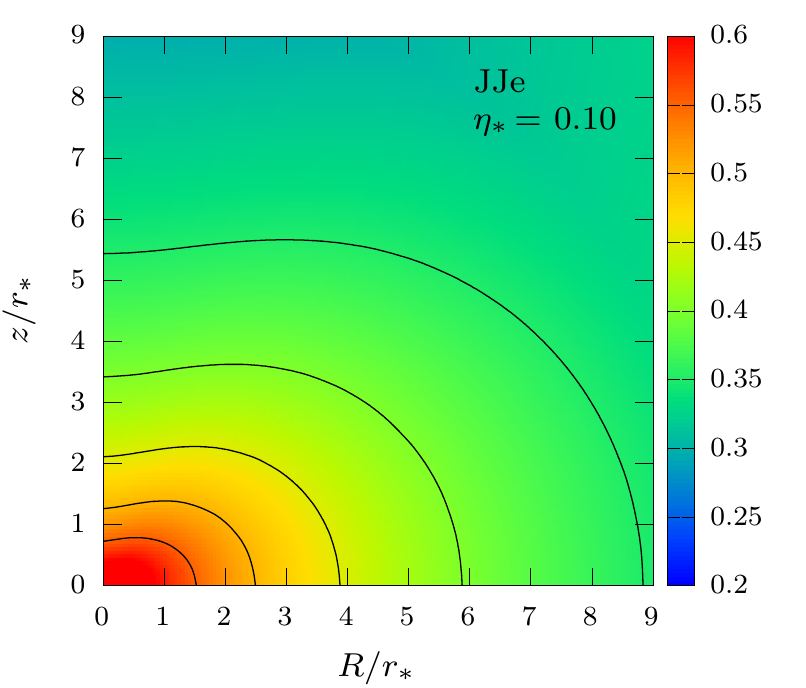}
  \hspace{2mm}
  \includegraphics[width=0.48\linewidth]{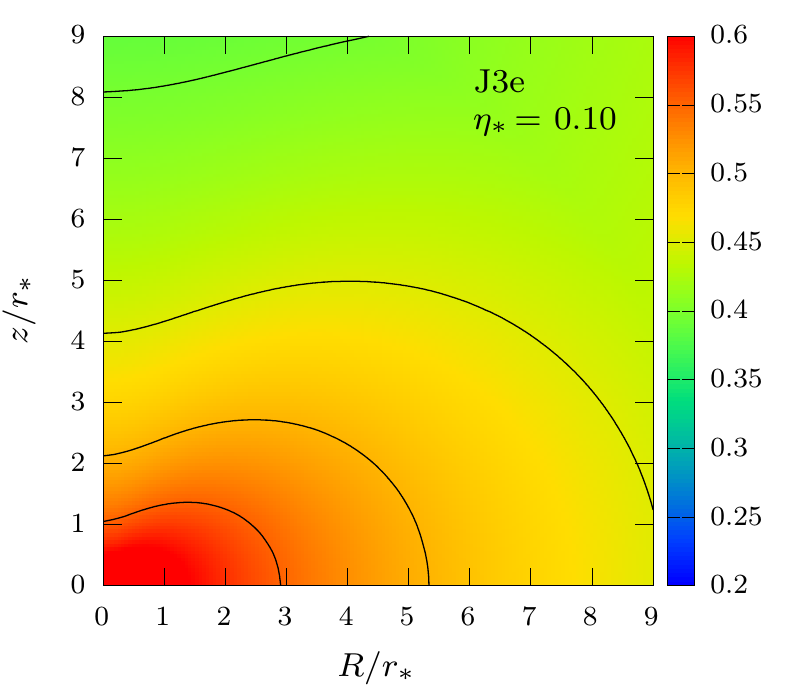}\vspace{2mm}
  \\
  \includegraphics[width=0.48\linewidth]{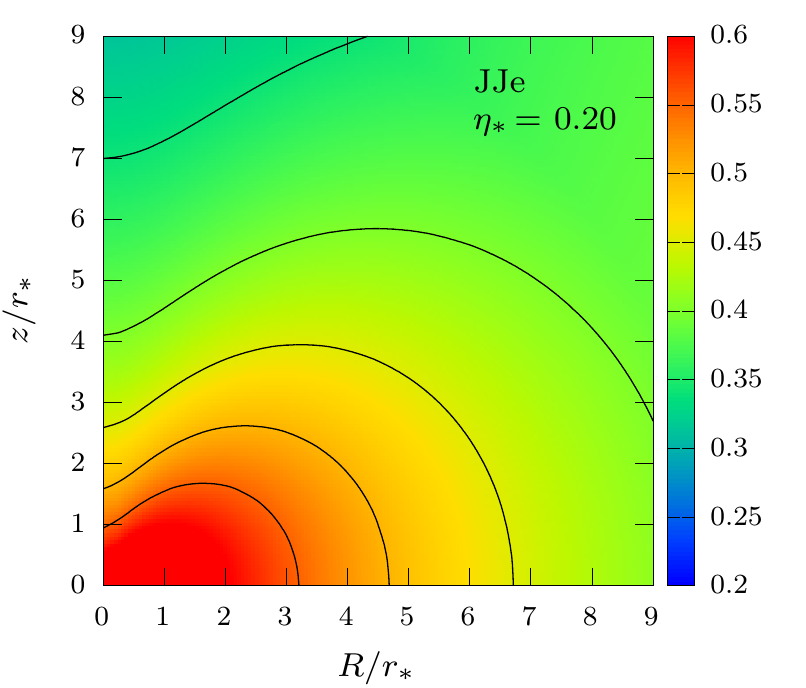}
  \hspace{2mm}
  \includegraphics[width=0.48\linewidth]{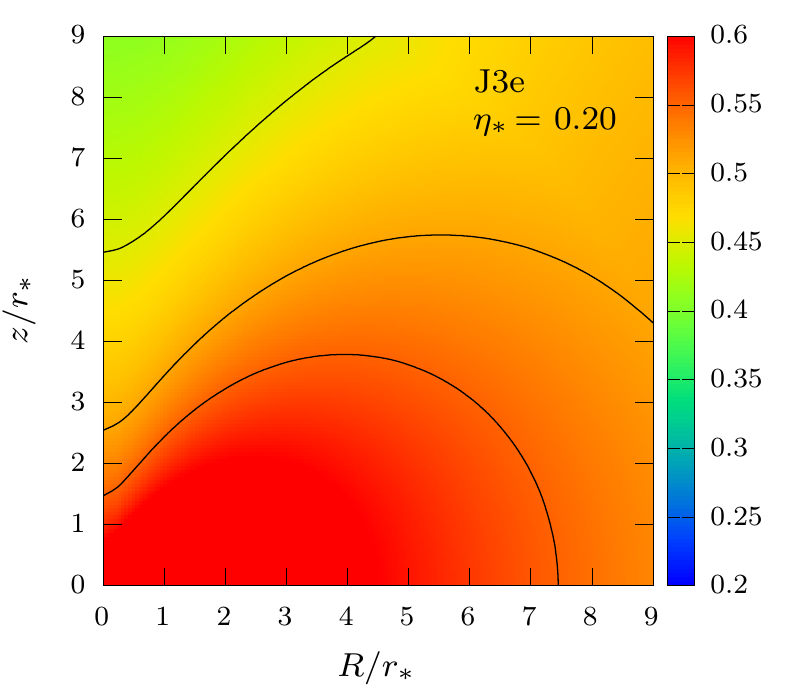}
  \vspace{0.5mm}
  \caption{Maps of $\sigphi(R,z)$, normalized to $\sqrt{\Psin}$, for the same minimum halo 
           JJe (left panels) and J3e (right panels) models of Fig. \protect\ref{fig:sigmas}, in 
           absence of rotation ($k=0$). The contour lines correspond to values that, 
           starting from $0.55$, decrease with step $0.05$ from inside to outside.}
  \label{fig:sigmaphi}
  \end{figure*}

  \begin{figure*}
  \includegraphics[width=0.48\linewidth]{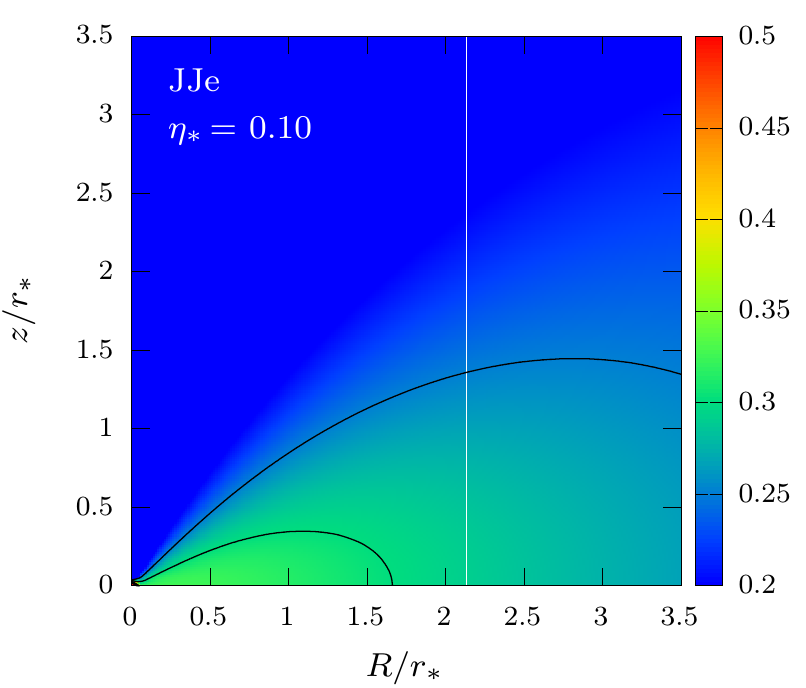}
  \hspace{4mm}
  \includegraphics[width=0.48\linewidth]{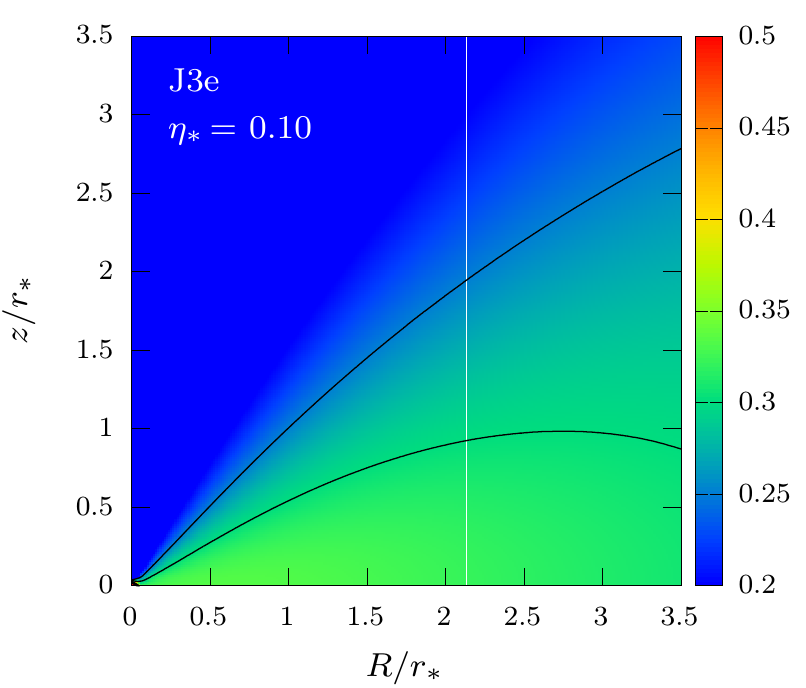}\vspace{2mm}
  \\
  \includegraphics[width=0.48\linewidth]{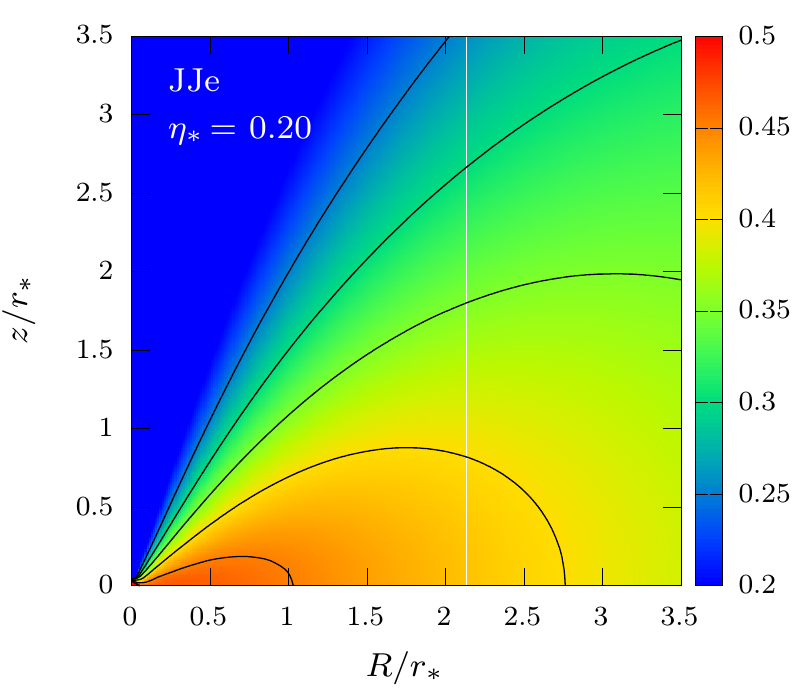}
  \hspace{4mm}
  \includegraphics[width=0.48\linewidth]{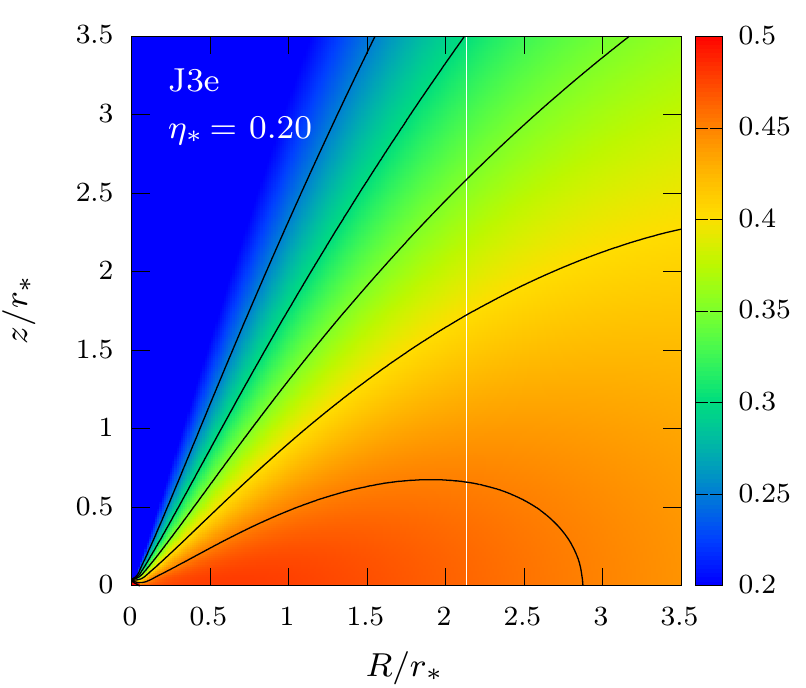}
  \vspace{0.5mm}
  \caption{Maps of $\vphib(R,z)$, normalized to $\sqrt{\Psin}$, for the same minimum halo JJe
           (left panels) and J3e (right panels) models of Fig. \protect\ref{fig:sigmas}, in 
           the isotropic ($k=1$) case. The contour lines correspond to values that, starting
           from $0.25$, increase with step $0.05$ from outside to inside. Note the different
           ranges of colour and axes values with respect to Fig. \protect\ref{fig:sigmas}
           and Fig. \protect\ref{fig:sigmaphi}.}
  \label{fig:vphim}
  \end{figure*}

  \section{The solution}\label{sec:Solution}

  First, we provide the solution of the vertical Jeans equation, retaining
  for simplicity only first order terms in the flattenings, as discussed 
  in detail in the previous Section. In full generality we have
  \begin{equation}
        \sigs^2=\sigsb^2+\sigsg^2,
  \end{equation}
  where $\sigsb$ and $\sigsg$ represent the contribution of the central
  BH and of the galaxy potential to the stellar velocity dispersion
  $\sigs$. From the expansion of $\rhos$ it follows that the 
  velocity dispersion profile due to the BH is given by
  \begin{equation}
        \rhos\sigsb^2=\rhon\Psin\mu\hspace{0.2mm}
                      \Big[
                      A(s)+\etas B(s)+\etas\Rtil^2 C(s)
                      \Big]\hspace{-0.1mm}, 
  \label{eq:ABC}
  \end{equation}
  where $A(s)$, $B(s)$, and $C(s)$ are obviously the same for the JJe and 
  J3e models, and are given in Appendix \ref{app:ABCDEFGH}. The formualae 
  of the contribution of $\Psig$ to the stellar velocity dispersion are 
  more complicated, depending also on the non-spherical component of the 
  galactic potential. At the linear order in the flattenings we have
  \begin{equation}
         \begin{aligned}
               \rhos\sigsg^2 &=\rhon\Psin\MR\hspace{0.2mm}
                               \Big[
                               D(s)+\etas E(s)\!\hspace{0.4mm}+\etas\Rtil^2 F(s)\\[5pt]
                             &+\etag G(s)+\etag\Rtil^2 H(s)
                               \Big]\hspace{-0.1mm},
         \end{aligned}
  \label{eq:DEFGH}
  \end{equation}
  and the functions from $D(s)$ to $H(s)$, for JJe and J3e models, are 
  given in Appendix C. Fig. \ref{fig:sigmas} shows a map 
  of $\sigs$ values in the meridional plane, for minimum halo models with 
  $\xi=5$, $\etag=0$, and $\mu=0.002$. Note the clear  elongation of the 
  curves with constant $\sigs$ along the $z$-axis; this behavior is 
  qualitatively explained by the oblate stellar density shape, to which 
  $\sigs$ must ``compensate'', in order for the product $\rhos\sigs^2$ to 
  be roughly spherical (see equation \ref{eq:Jeans_vert_sol}, where 
  $\PsiT$ is spherical).

  Second, we evaluate $\Dels=\Delsbh + \Delsg$ from the radial Jeans 
  equation, where $\Delsbh$ and $\Delsg$ are the contribution of the BH 
  and of the galaxy potential, respectively. At the linear order in the
  flattenings, by using the general considerations in Section 
  \ref{sec:radial}, and in particular equation \eqref{eq:commutator}, 
  remarkable identity holds
  \begin{equation}
        \rhos\Delsbh = 2\rhon\Psin\mu\hspace{0.3mm}\etas\Rtil^2 C(s),
  \label{eq:rhos_DeltaBH}
  \end{equation}
  where $C(s)$ is the same function appearing in equation \eqref{eq:ABC}. 
  As anticipated in Section \ref{subsec:vert}, it is possible to solve 
  analytically the full homoeoidal problem, but for simplicity here we limit 
  ourselves to the first order expansion. Thus, the galactic contribution
  to $\Dels$ is given by 
  \begin{equation}
        \rhos\Delsg=2\rhon\Psin\MR\hspace{0.2mm}\Rtil^2 
                    \Big[
                    \etas F(s)+\etag H(s)-\etag\tilde\rho_{*0}(s)\tPsigd(s)
                    \Big]\hspace{-0.1mm},
  \label{eq:rhos_Deltag}
  \end{equation}
  where $F(s)$ and $H(s)$ are the same functions in equation 
  \eqref{eq:DEFGH}, and $\tPsigd(s)$ is given in equations 
  \eqref{eq:tPsigi_JJe} and \eqref{eq:tPsigi_J3e} for JJe and J3e models. 
  Note that, as expected, the two contributions to $\rhos\Dels$ vanish for 
  $\etas=\etag=0$. 

  Fig. \ref{fig:sigmaphi} shows a map of $\sigphi$ values in the meridional 
  plane, for the same minimum halo models of the previous Fig. 
  \ref{fig:sigmas}, and in absence of net rotation (i.e., $k=0$). In both 
  Figs. \ref{fig:sigmas} and \ref{fig:sigmaphi} one can note the flatter 
  decline of $\sigs$ and $\sigphi$, moving outward from the center, for 
  J3e than for JJe models, due to the shallower DM distribution of J3e 
  models (see Fig. \ref{fig:iso_dens_JJe}).\\

  Fig. \ref{fig:vcVSvphimRz} shows $\vphib$ in the equatorial plane, 
  compared with $\vcirc(R\hspace{0.25mm})$, for {\it isotropic} ($k=1$) 
  minimum halo models, with $\xi=5$, $\etag=0$, and $\mu=0.002$, and two 
  shapes for the stellar density ($\etas=0.10$ and $0.30$). For the same 
  models, but $\etas=0.10$ and $0.20$, Fig. \ref{fig:vphim} shows the maps 
  of $\vphib$ in the meridional plane. In both Figs. \ref{fig:vcVSvphimRz} 
  and \ref{fig:vphim} the values of $\vphib$ keep larger for the J3e model 
  than for the JJe one, at the same distance from the galactic center, 
  again due to the shallower DM distribution (see Fig. 
  \ref{fig:iso_dens_JJe}).

\begin{figure*}
  \includegraphics[width=0.493\linewidth]{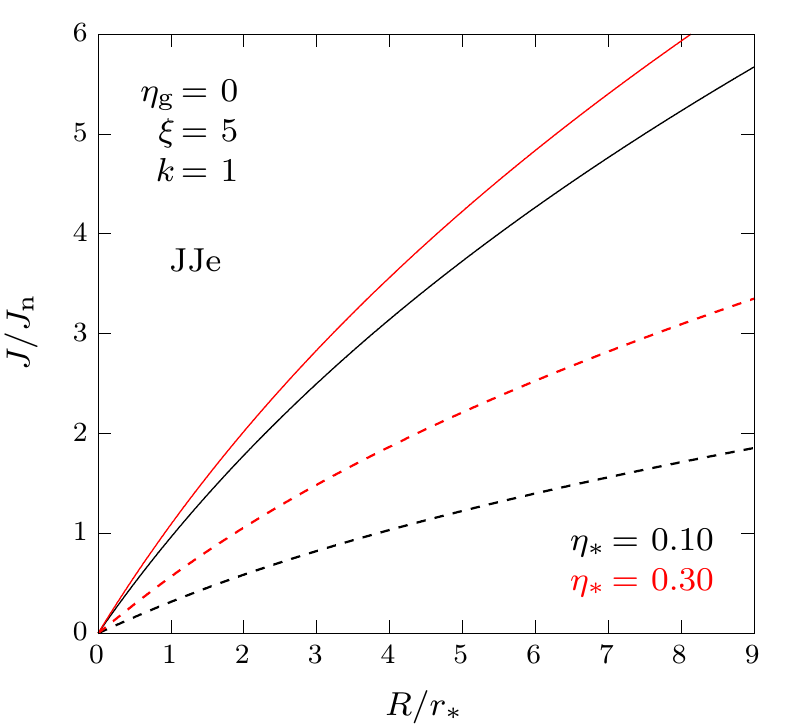}
  \hspace{0.8mm}
  \includegraphics[width=0.493\linewidth]{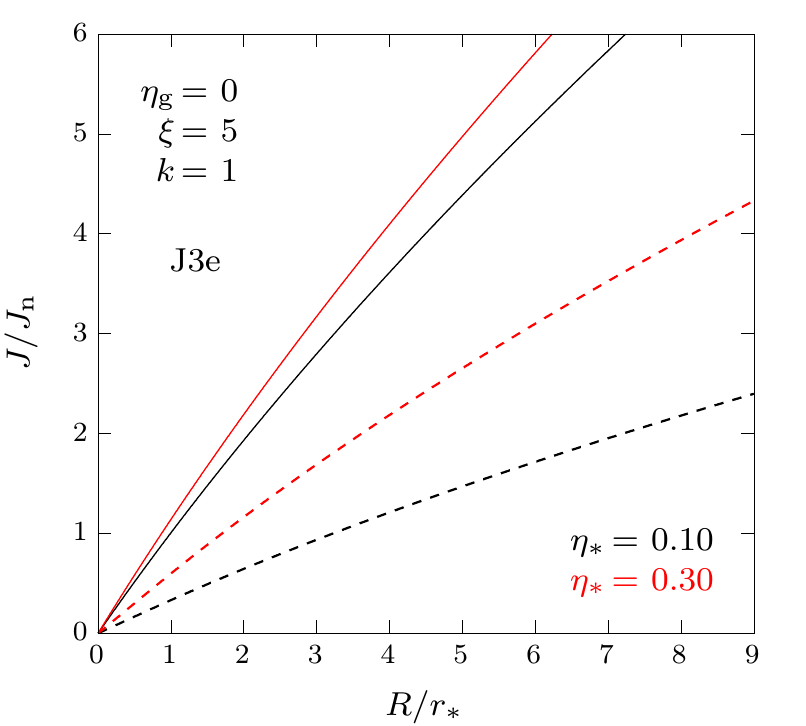}
  \vspace{-2.5mm}
  \caption{Radial trends of $J_{\rm c}(R\hspace{0.25mm})=R\,\vcirc(R\hspace{0.25mm})$ (solid 
           lines) and $J_z(R,0)=R\,\vphib\hspace{0.4mm}(R,0)$ (dashed lines), normalized to 
           $J_{\rm n} = \rs\sqrt{\Psin}=\sqrt{G\Ms\rs}$, for the same isotropic minimum halo
           JJe (left) and J3e (right) models of Fig. \protect\ref{fig:vcVSvphimRz}.}
  \label{fig:JcVSJz}
  \end{figure*}

\begin{figure*}
  \includegraphics[width=0.494\linewidth]{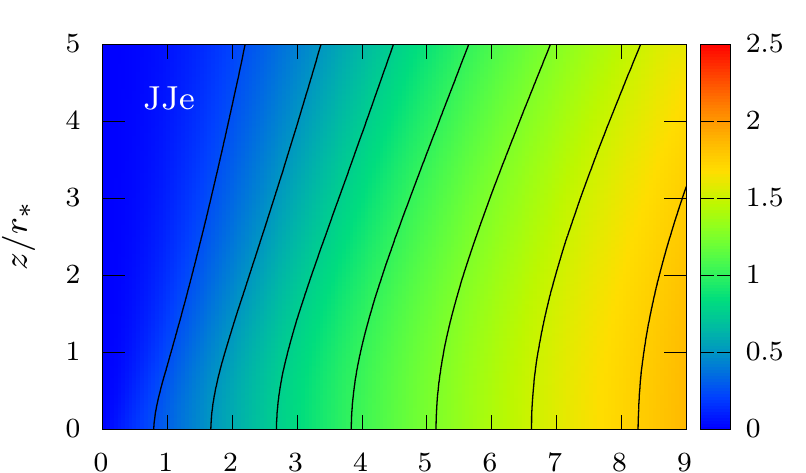}
  \hspace{0.83mm}
  \includegraphics[width=0.494\linewidth]{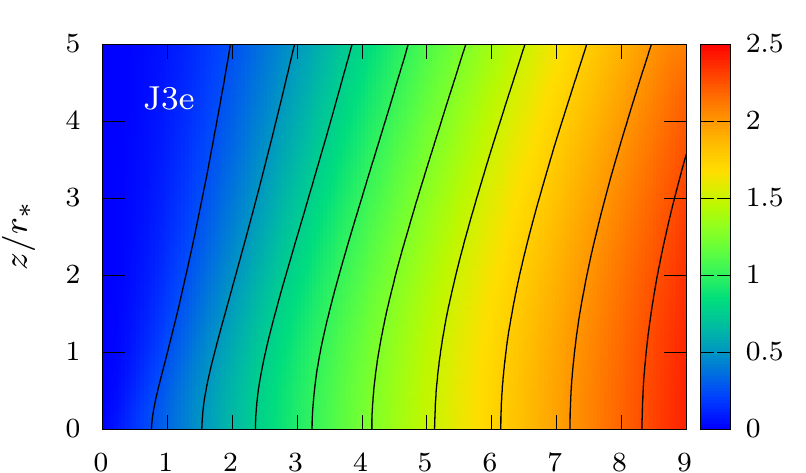}\\
  \vspace{-0.2mm}
  \includegraphics[width=0.494\linewidth]{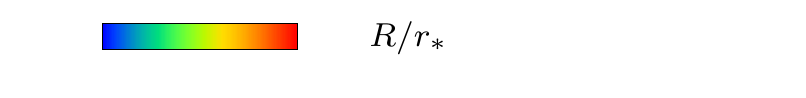}\hspace{1.45mm}
  \includegraphics[width=0.494\linewidth]{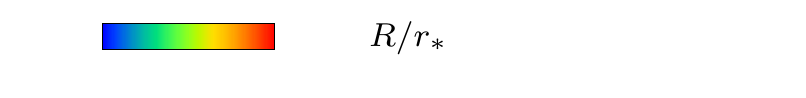}
  \vspace{-6.5mm}
  \caption{Maps of $J_z(R,z)=R\,\vphib(R,z)$, normalized to $J_{\rm n}$, for the same 
           isotropic  minimum halo models in the two top panels of Fig. \protect\ref{fig:vphim}. The 
           black contour lines correspond to values that, starting from $0.25$, increase with 
           step $0.25$ from inside to outside. The bottom colour bars show 
           $J_{\rm c}(R\hspace{0.25mm})=R\,\vcirc(R\hspace{0.25mm})$, normalized to 
           $J_{\rm n}$, up to a value of $R$ whereby $J_{\rm c}(R\hspace{0.25mm})/J_{\rm n}=2.5$ 
           (the discussion at the end of Sect. \protect\ref{sec:Solution} illustrates the use of such 
           a bar).}
  \label{fig:Jz_Jcbar}
  \end{figure*}

  Finally, we present in Fig. \ref{fig:JcVSJz} the trends on the equatorial 
  plane of two angular momenta per unit mass, 
  $\Jc(R\hspace{0.25mm})=R\hspace{0.25mm}\vcirc(R\hspace{0.25mm})$ and 
  $J_z(R\hspace{0.25mm})=R\hspace{0.35mm}\vphib(R,0)$, for the same models 
  of the Fig. \ref{fig:vcVSvphimRz}. The circular orbit with velocity 
  $\vcirc$ corresponds to the minimum energy, and thus this figure gives an 
  idea of the radius $R$ where a unit mass ends up in the equatorial plane, 
  after dissipating the maximum possible of its energy, while conserving 
  angular momentum. The practical case here is that of a parcel of gas with 
  a specific angular momentum $J_z(R,z)$, that falls on the equatorial plane 
  at some $R$, and then moves inward until it reaches the minimum $\Rin$ 
  corresponding to $\Jc(\Rin)=J_z(R,0)$. If the gas origin is in stars close 
  to the equatorial plane, and the gas inherits the $J_z(R,0)$ of its parent 
  stars, then it moves inward crossing a radial interval $R-\Rin$ given by 
  the condition $\Jc(\Rin)\simeq J_z(R,0)$; this interval can be derived 
  directly from Fig. \ref{fig:JcVSJz}. Fig. \ref{fig:Jz_Jcbar} further 
  illustrates these points: the maps of $J_z(R,z)$ give an idea of where gas 
  may end up if falling on the equatorial plane; the colour bars below the 
  maps show the $\Jc(R\hspace{0.25mm})$ values for a short $R$-range, and 
  allow to link the various $J_z(R,0)$ to the minimum radii $\Rin$ that the 
  gas reaches through motions at constant $J_z$ but dissipating energy, 
  while on the equatorial plane. For the models in the Figures, gas from 
  the bulk of the galaxies ($\simeq 7.5\hspace{0.3mm}\rs$) is expected to 
  end up within ``disks'' of just $R \la 2\hspace{0.2mm}\rs$. 

  \subsection{Asymptotic behavior} \label{subsec:asymp_3D}

  A more quantitative analysis of the effects of the model parameters on 
  the dynamical properties of the stellar component is provided by the 
  asymptotic expansion of the solutions near the center and at large radii. 

  Near the center (i.e., for $s\to 0$), from Taylor expansion with 
  $\Rtil=s\,\sin\theta$, the asymptotic behaviour of $\rhos\sigs^2$ and 
  $\rhos\Dels$ for JJe models coincides with that of J3e models. By expanding 
  up to the dominant term of the galaxy contribution we find
  \begin{equation}
        \begin{aligned}
              \frac{\rhos\sigs^2}{\rhon\Psin} &\sim 
                                               \mu\hspace{-0.2mm}\left[\frac{5+\etas(1-6\cos^2\!\theta)}{15s^3}-\frac{2+\etas\!\hspace{0.2mm}\sin^2\!\theta}{2s^2}\right]\\[8pt]
                                              &+\MR\,\frac{3(1-\etas\!\hspace{0.2mm}\cos^2\!\theta)+\etag(1+\sin^2\!\theta)}{6\xi s^2},
        \end{aligned}
  \label{eq:rhos_sigs_center}
  \end{equation}
  and
  \begin{equation}
        \frac{\rhos\Dels}{\rhon\Psin}\sim
        \mu\hspace{-0.2mm}\left(\frac{4}{5s^3}-\frac{1}{s^2}\right)\!\etas\!\hspace{0.2mm}\sin^2\!\theta
+\MR\,\frac{(3\etas\!-\etag)\!\hspace{0.3mm}\sin^2\!\theta}{3\xi s^2}.
  \label{eq:rhos_dels_center}
  \end{equation}

  At large radii (i.e., when $s\to\infty$) we have
  \begin{equation}
        \frac{\rhos\sigs^2}{\rhon\Psin} \sim \frac{7-\etas(1+20\cos^2\!\theta)}{35 s^5} \times 
        \begin{cases}
              \hspace{0.05cm}\displaystyle{\MR+\mu}, \hspace{0.6cm} (\JJe),
              \\[6pt]
              \hspace{0.05cm}\displaystyle{\MR\ln s}, \hspace{0.666cm} (\Jte),
        \end{cases}
  \label{eq:rhos_sigs_inf}
  \end{equation}
  and
  \begin{equation}
        \frac{\rhos\Dels}{\rhon\Psin} \sim \frac{8\hspace{0.2mm}\etas\!\hspace{0.2mm}\sin^2\!\theta}{7s^5} \times 
        \begin{cases}
              \hspace{0.05cm}\displaystyle{\MR+\mu} \hspace{0.6cm} (\JJe), 
              \\[6pt]
              \hspace{0.05cm}\displaystyle{\MR\ln s} \hspace{0.666cm} (\Jte).
        \end{cases}
  \end{equation}

  The trends above suggest three comments. First, equations 
  \eqref{eq:rhos_sigs_center} and \eqref{eq:rhos_sigs_inf} coincide, for 
  $\etas=\etag=0$, with the analogous formulae in CZ18 and CMP19 for the 
  fully isotropic case, as expected. Second, note how the BH mass appears 
  in equation \eqref{eq:rhos_sigs_inf} for JJe models, due to the total 
  finite mass, so that the $\sigs$ is dominated by the monopole term of 
  $\Psig$. Of course, the presence of $\mu$ is totally irrelevant for any 
  practical application. For the same reason, $\mu$ does not appear in 
  the case of the J3e models, which have an infinite mass. Third, the 
  present models exhibit a peculiar behaviour, i.e., near the center 
  the velocity dispersion in the non-spherical case for $\mu=0$ is 
  {\it finite but discontinuous}: approaching the center along different 
  $\theta$ directions, one determines different values of the velocity 
  dispersion $\sigs$. 
  This results from the non-spherical shape of $\rhos$, the central 
  slope of $\rhos$, and the gravitational potential entering the Jeans 
  equations. For example, by using equation (C.3) in Ciotti \& Bertin (2005; 
  see also equation A.4 in Riciputi et al. 2005), it is easy to prove that 
  in the self-gravitating case, the central velocity dispersion for the density 
  profile $1/m^\gamma$ is zero for $0<\gamma<2$, finite discountinuous (as the 
  models in this paper) for $\gamma=2$, and infinite for $\gamma>2$. Instead, 
  $\sigs \to \infty$ for generic $\gamma$ in presence of a central BH, while 
  $\sigs$ is finite discontinuous for generic values of $\gamma$ if the 
  ellipsoid is embedded in the potential of the Singular Isothermal Sphere 
  (Ciotti 2021).
  
  \subsection{Asymmetric Drift}

  Asymmetric drift (see e.g. BT08 for a definition) plays some role in 
  the phenomenon of radial gas flows in rotating systems (see Smet 2015). 
  Its computation presents no difficulties in the framework of homoeoidal 
  expansion of our models, thus here we present the basic formulae. We 
  recall that the asymmetric drift (hereafter, $\AD$) in the equatorial 
  plane is defined as $\AD = \vcirc-\,\vphib$. In turn, we define the 
  function $\Deltac\equiv\vcirc^2-\,\vphimq$, which is of easier evaluation 
  in analytical studies; indeed, for moderate values of the asymmetric 
  drift, one has $\AD\simeq\Deltac/(2\vcirc)$. By virtue of equations 
  \eqref{eq:Jeans_rad} and \eqref{eq:vphimedio_Satoh} we readily have
  \begin{equation} 
        \rhos\Deltac= 
        \big(1-k^2\hspace{0.1mm}\big)\rhos\Dels\!\hspace{0.2mm}-\hspace{0.2mm}R\,\hspace{0.1mm}\frac{\partial \rhos\sigs^2}{\partial R}, 
        \qquad\,\,\, (z=0). 
  \label{eq:rhos_Dc_general}
  \end{equation}
  For example, in the isotropic case ($k=1$), by using equations 
  \eqref{eq:ABC} and \eqref{eq:DEFGH}, a Taylor expansion near the center 
  shows that, at the linear order in the flattenings,
  \begin{equation}
        \frac{\Deltac}{\Psin} \sim \mu\!\left(\frac{5-4\hspace{0.1mm}\etas}{5\Rtil}-2+\etas\!\right)\!
+\MR\,\frac{3\hspace{0.1mm}(1-\etas)+2\hspace{0.1mm}\etag}{3\xi}.
  \label{eq:AD_center}
  \end{equation}
  In presence of a dominant central BH, one finds $\Deltac\propto R^{-1}$ 
  with a nowhere negative proportionality constant. When $\mu=0$, instead, 
  $\Deltac$ reduces to a constant value, but also in this case 
  $\Deltac \geq 0$ for reasonable values of $\etas$ (which always cannot 
  exceed $1/3$) and $\etag$ (which cannot exceed $1/3$ for JJe models, and 
  $1/2$ for J3e models).  

  In the external regions (i.e., when $R \to \infty$), instead, at the 
  leading order we find
  \begin{equation}
        \frac{\Deltac}{\Psin} \sim \frac{7-8\hspace{0.1mm}\etas}{7\Rtil} \times 
        \begin{cases}
              \hspace{0.05cm}\displaystyle{\MR+\mu}, \hspace{0.634cm} (\JJe),
              \\[8pt]
              \hspace{0.05cm}\displaystyle{\MR\ln\Rtil}, \hspace{0.6cm} (\Jte).
        \end{cases}
  \label{eq:AD_inf}
  \end{equation}
  By considering the dominant term of the two previous equations, 
  $\Deltac \geq 0$ for reasonable values of $\etas$, as expected. For 
  general values of $k$, it is sufficient to add in equation 
  \eqref{eq:rhos_Dc_general} the expression for $\rhos\Dels$ (when $z=0$) 
  derived in the previous Section.

  \section{Projected Dynamics}\label{sec:Proj}

  The projection of a galaxy model on the plane of the sky is an 
  important step in the model construction, needed in order to 
  determine the observational properties of the model itself. In 
  this paper we deal with axisymmetric models, so we need to specify 
  just a single angle $i$ that gives the direction of the line 
  of sight (hereafter, l.o.s.) to the observer. Moreover, the simple 
  functional form of the density and of the intrinsic kinematical 
  fields in the homoeoidal framework leads to further simplifications. 

  Let our models be described in a Cartesian inertial frame of 
  reference $S_0$, with coordinates $(x,y,z)$. In addition, consider 
  a second orthogonal reference system $S'$, with coordinates 
  $(X,Y,Z)$, and same origin as $S_0$. Due to axisymmetry, 
  and without loss of generality, the relation between the two sets of 
  coordinates is given by
  \begin{equation}
        \begin{pmatrix}
               x \\
               y \\
               z
        \end{pmatrix}=
        \begin{pmatrix}
               \cos i  &  0  &  \sin i  \\
                   0       &  1  &      0       \\
              -\sin i  &  0  &   \cos i \hspace{0.5mm}
       \end{pmatrix}\!\!
       \begin{pmatrix}
              X \\
              Y \\
              Z
       \end{pmatrix}\!\hspace{-0.2mm},
  \label{eq:coord_trans}
  \end{equation}
  with the l.o.s. being directed along $Z$. With this choice, the 
  (transpose) of the unit vector ${\bf n}$, from $S_0$ to the observer, 
  is ${\bf n}=(\sin i,0,\cos i)$. Moreover, we consider 
  a counter clockwise rotation of an angle $i$ around the $y$-axis, 
  coincident with the $Y$-axis of the observer. In particular, for 
  $i=0$, corresponding to the so-called {\it face-on} projection, 
  the l.o.s. coincides with the $z$-axis, and $(X,Y)=(x,y)$; for 
  $i=\upi/2$, corresponding to the so-called {\it edge-on} projection, 
  the l.o.s. coincides with the $x$-axis, and $(X,Y)=(-z,y)$. In 
  any case, $Y$ is aligned with the major axis of the projection. 
  The distance from the center of the image is indicated by 
  $R=\sqrt{X^2+Y^2}$.

  Accordingly, the projection of the stellar density is given by
  \begin{equation}
        \Sigmas\hspace{0.2mm}\equiv\int_{-\infty}^{\infty}\!\rhos dZ,
  \label{eq:Sigmas_proj}
  \end{equation}
  where the function $\rhos$ is expressed from equation 
  \eqref{eq:coord_trans} in terms of vector ${\bf X}$ and angle $i$.

  The projection of the component along $\nv$ of the ordered velocity, 
  called {\it l.o.s. streaming velocity field} and indicated with $\vlos$, 
  is given by
  \begin{equation}
        \Sigmas\vlos\hspace{0.2mm}\equiv
        \int_{-\infty}^{\infty}\!\rhos\vmedio\cdot\nv\hspace{0.5mm}dZ,
  \label{eq:vlos}
  \end{equation}
  where the overline represents the mean over the phace-space; again, as in 
  equation \eqref{eq:Sigmas_proj}, the left hand side of equation 
  \eqref{eq:vlos} depends on $X$, $Y$, and $i$. Note that, 
  since the streaming motion of stars occurs only in the azimuthal direction, 
  we have $\vmedio\cdot\nv=-\,\vphib\sin i\sin\varphi$; as a consequence, 
  $\vphib > 0$ when the galaxy rotates in an anti-clockwise sense. In the 
  limit case of a face-on projection, we have no streaming motion. 

  The projection of the squared velocity reads
  \begin{equation}
        \Sigmas\vp^2\hspace{0.2mm}\equiv
        \int_{-\infty}^{\infty}\!\rhos\overline{(\vv\cdot\nv)^2}\hspace{0.5mm}dZ.
  \label{eq:vp}
  \end{equation}
  By introducing the velocity dispersion tensor $\sigij^2$, and expanding 
  equation \eqref{eq:vp}, we have $\vp^2=\sigp^2+\Vp^2$, where\footnote{The 
  quantity $\vp$ coincides with $V_{\rm rms}$ in Cappellari et al. (2013).}
  \begin{equation}
        \Sigmas\sigp^2\hspace{0.2mm}\equiv\int_{-\infty}^{\infty}\!\rhos\sigij^2 n_i n_j dZ, 
        \qquad
        \Sigmas\Vp^2\hspace{0.2mm}\equiv
        \int_{-\infty}^{\infty}\!\rhos(\vmedio\cdot\nv)^2\hspace{0.1mm}dZ.
  \end{equation}
  Note that, if a net rotation is present, the projected velocity dispersion 
  $\sigp^2$ is not the observed velocity dispersion $\siglos$; the l.o.s. 
  velocity dispersion, which is related to the broadening of the spectral 
  lines, is instead given by
  \begin{equation}
        \Sigmas\siglos^2\equiv
        \int_{-\infty}^{\infty}\!\rhos\overline{(\vv\cdot\nv-\vlos)^2}\hspace{0.5mm}dZ.
  \label{eq:siglos}
  \end{equation}
  Finally, by combining equations \eqref{eq:vlos}, \eqref{eq:vp}, and 
  \eqref{eq:siglos}, one has
  \begin{equation}
        \siglos^2=\sigp^2+\Vp^2-\vlos^2.
  \label{eq:sigma_los}
  \end{equation}
  In a face-on projection, as $\nv$ is always perpendicular to $\vmedio$, 
  the previous equation reduces to $\siglos=\sigp$. Moreover, by adopting 
  the Satoh $k$-decomposition, it is easy to show that the coefficient $k$ 
  appears only in $\vlos$; indeed, using equations \eqref{eq:vphimedio_Satoh} 
  and \eqref{eq:sigmaphi_Satoh} one has
  \begin{equation}
        \Sigmas\vp^2=\int_{-\infty}^{\infty}\!
        \rhos\big(\sigs^2+\Dels\!\hspace{0.3mm}\sin^2\hspace{-0.3mm}i\sin^2\!\varphi\big)dZ,
  \end{equation}
  and, in case of a constant $k$,
  \begin{equation}
        \Sigmas\vlos=-\,k\int_{-\infty}^{\infty}\!\rhos\sqrt{\Dels}\!\hspace{0.5mm}\sin i\sin\varphi\,dZ.
  \end{equation}\\
  In the more general case of a coordinate-dependent $k$, this parameter 
  would appear inside the integral. Of course, since $\Dels$ vanishes when 
  $\etas=\etag=0$, spherical models present no streaming motion.

  \subsection{Asymptotic behaviour}

  The projection integrals in the previous Section in general may be 
  performed only numerically (see Caravita et al. 2020). Here we focus on 
  the asymptotic expansion of the projected fields at small and large 
  radii, amenable to analytical treatment. 

  From Taylor expansion, the asymptotic behaviour of the projected stellar 
  density, for {\it both} JJe and J3e models, can be written as
  \begin{equation}
        \frac{\Sigmas}{\Sign}\sim\upi
        \begin{cases}
              \hspace{0.05cm}\displaystyle{\frac{\Rtil^2\hspace{-0.4mm}+\etas\tY^2\hspace{-0.2mm}\sin^2\hspace{-0.3mm}i}{\Rtil^3}}, \hspace{1.8cm} \Rtil \to 0,
              \\[13pt]
              \hspace{0.05cm}\displaystyle{\frac{\Rtil^2\hspace{-0.4mm}+\etas(\hspace{0.1mm}\tY^2\hspace{-0.4mm}-2\tX^2)\hspace{-0.2mm}\sin^2\hspace{-0.3mm}i}{2\Rtil^5}}, \hspace{0.578cm} \Rtil \to \infty,
        \end{cases}
  \end{equation}
  where $\Sign \equiv \Ms/(4\upi\rs^2)$, and the tilde indicates the normalization
  with respect to $\rs$.

  In the face-on case, for which $X=x$ and $Y=y$, we have $\sigij^2 n_i n_j=\sigs^2$. Again, 
  from the asymptotic analysis of $\rhos\sigs^2$ (see Section 
  \ref{subsec:asymp_3D}), the behaviour of $\siglos$ in the central region 
  is the same for both JJe and J3e models, and it reads
  \begin{equation}
        \frac{\Sigmas\siglos^2}{\Sign\Psin} \sim
        \mu\hspace{-0.2mm}\left[\frac{2\hspace{0.1mm}(5-\etas)}{15\Rtil^2}-\frac{\upi(4+\etas)}{4\Rtil}\right]\!+
\MR\,\frac{\upi\hspace{0.1mm}(2-\etas+\etag)}{4\xi\Rtil}.
  \label{eq:Sigma_siglos2_FOcenter}
  \end{equation}
  Notice that, in absence of the central BH, $\siglos^2$ reduces to a 
  nowhere negative constant value for acceptable values of $\etas$ and 
  $\etag$.
  Very far from the center, instead, at the leading order we have
  \begin{equation}
        \frac{\Sigmas\siglos^2}{\Sign\Psin} \sim \frac{4(7-5\hspace{0.1mm}\etas)}{105\Rtil^4} \times
        \begin{cases}
              \hspace{0.05cm}\displaystyle{\MR+\mu}, \hspace{0.65cm} (\JJe),
              \\[6pt]
              \hspace{0.05cm}\displaystyle{\MR\ln\Rtil}, \hspace{0.616cm} (\Jte).
        \end{cases}
  \label{eq:Sigma_siglos2_FOinfinity}
  \end{equation}

  In the edge-on case, in which $X=-z$ and $Y=y$, the quantities 
  $\Vp$ and $\vlos$ are not zero. By considering only the leading order terms 
  of equations \eqref{eq:rhos_sigs_center} and \eqref{eq:rhos_dels_center}, 
  and limiting to the lowest in the flattenings, some careful algebra shows 
  that, for $\Rtil \to 0$,
  \begin{equation}
        \frac{\Sigmas\siglos^2}{\Sign\Psin} \sim 
        \begin{cases}
             \hspace{0.05cm}\displaystyle{\MR\,\frac{\ag(\tX,\tY)}{\Rtil}}, \hspace{0.9cm} (\mu=0),
             \\[11pt]
             \hspace{0.05cm}\displaystyle{\mu\,\frac{\aBH(\tX,\tY)}{\Rtil^2}}, \hspace{0.745cm} (\mu \ne 0),
        \end{cases}
  \label{eq:a&b(lambda)}
  \end{equation}
  and
  \begin{equation}
        \frac{\Sigmas\vlos}{\Sign\sqrt{\Psin}} \sim -\hspace{0.4mm}k 
        \begin{cases}
              \displaystyle{\sqrt{\MR\,\frac{3\hspace{0.2mm}\etas\hspace{-0.6mm}-\etag}{3\xi}}\hspace{0.4mm}\frac{2\hspace{0.1mm}\tY}{\Rtil^2}}, \hspace{1.2cm} (\mu=0),
              \\[12pt]
              \displaystyle{\sqrt{\frac{\mu\etas}{5}}\hspace{0.4mm}B\hspace{-0.6mm}\left(\frac{1}{2},\frac{5}{4}\right)\!\frac{2\hspace{0.1mm}\tY}{\Rtil^{5/2}}}, \hspace{0.575cm} (\mu \ne 0),
        \end{cases}
  \end{equation}
  where $B(p,q)$ is Euler's complete Beta function. In the external regions, instead,
  \begin{equation}
        \frac{\Sigmas\siglos^2}{\Sign\Psin} \sim \frac{4\hspace{0.3mm}b(\tX,\tY)}{7\Rtil^4} \times
        \begin{cases}
              \hspace{0.05cm}\displaystyle{\MR+\mu}, \hspace{0.65cm} (\JJe),
              \\[6pt]
              \hspace{0.05cm}\displaystyle{\MR\ln\Rtil}, \hspace{0.616cm} (\Jte),
        \end{cases}
  \label{eq:c(lambda)}
  \end{equation}
  and
  \begin{equation}
        \frac{\Sigmas\vlos}{\Sign\sqrt{\Psin}} \sim 
  -\hspace{0.4mm}k\hspace{0.3mm}\sqrt{\frac{2\hspace{0.2mm}\etas}{7}}\hspace{0.4mm}B\hspace{-0.6mm}\left(\frac{1}{2},\frac{9}{4}\right)\!\frac{2\hspace{0.1mm}\tY}{\Rtil^{9/2}}\times
        \begin{cases}
              \hspace{0.05cm}\displaystyle{\sqrt{\MR+\mu}}, \hspace{0.4cm} (\JJe),
              \\[6pt]
              \hspace{0.05cm}\displaystyle{\sqrt{\MR\ln\Rtil}}, \hspace{0.366cm} (\Jte);
        \end{cases}
  \end{equation}
  the functions $\ag$, $\aBH$, and $b$, are given in Appendix \ref{app:sigmalos_edgeon}.
  As expected, 
  equations \eqref{eq:Sigma_siglos2_FOcenter}, \eqref{eq:Sigma_siglos2_FOinfinity}, 
  \eqref{eq:a&b(lambda)} and \eqref{eq:c(lambda)} reduce, for $\etas=\etag=0$, 
  to the analogous formulae given in CZ18 and CMP19 for the spherical fully 
  isotropic case.
  
  \section{The Virial Theorem}\label{sec:VT}

  The Virial Theorem (hereafter, VT) provides important information 
  about the global energetics of a galaxy model. Here we focus on the 
  VT of the stellar components:
  \begin{equation}
         2\Ks = -\,\Ws \equiv -\,\Wg -\Wbh,
  \end{equation}
  where
  \begin{equation}
        \Ks=
           \frac{1}{2}\int\hspace{-0.2mm}\rhos\big(\hspace{0.1mm}2\hspace{0.1mm}\sigs^2+\overline{\vphi^2}\hspace{0.5mm}\big)d^3\hspace{0.15mm}\xv
           =\frac{1}{2}\int\!\rhos\big(3\hspace{0.1mm}\sigs^2+\hspace{0.15mm}\Dels\!\hspace{0.25mm}\big)d^3\xv
  \end{equation} 
  is the {\it total} kinetic energy of the stars, 
  \begin{equation}
        \Wg= \int \!\rhos\xv\cdot\frac{\partial\Psig}{\partial\xv}\,d^3\xv
  \label{eq:Wsg}
  \end{equation}
  is the interaction energy of the stars with the gravitational field 
  of the galaxy (stars plus DM), and finally
  \begin{equation}
        \Wbh=\int\!\rhos\xv\cdot\frac{\partial\Psibh}{\partial\xv}\,d^3\xv
            =\Ubh,
  \end{equation}
  where $\Ubh=-\int \rhos \Psibh d^3\xv$ is the gravitational energy of 
  the stars due to the central BH. Note that for a Jaffe density 
  distribution, $\Wbh$ diverges near the origin, so that also the volume 
  integral of $\rhos\sigsb^2$ diverges, as can be verified by direct 
  integration of equation \eqref{eq:ABC}.

  In the framework of homoeoidal expansion the integrand in equation 
  \eqref{eq:Wsg} can be expressed in simple form. Indeed, with some 
  work\footnote{From equation \eqref{eq:Psig_exp}, simple algebra shows 
  that
  \begin{equation*}
        \xv\cdot\frac{\partial\Psig}{\partial\xv} = \Psin\MR\hspace{0.1mm}
        \bigg[\bigg(\frac{d\tPsigz}{ds}+\etag\frac{d\tPsigu}{ds}\bigg)\hspace{-0.1mm}s + \etag\Rtil^2\!\hspace{0.2mm}\bigg(s\hspace{0.2mm}\frac{d\tPsigd}{ds}+2\hspace{0.1mm}\tPsigd\bigg)\hspace{-0.1mm}
        \bigg].
  \label{eq:xpos_dot_gradPsig}
  \end{equation*}},
  after transforming variables to spherical coordinates, an integration 
  over the solid angle shows that, limiting to linear terms in the 
  flattenings, 
  \begin{equation}
        \Wg = -\,\Un\MR\times\left(w_0+\etas w_1+ \etag w_2\right)\!,
  \label{eq:w012}
  \end{equation}
  where $\Un \equiv \Ms\Psin$, and
  \begin{equation}
         w_i =
         \begin{cases} 
               \hspace{0.05cm}\displaystyle{-\!\int_0^{\infty}\!\tilde\rho_{*0}\hspace{0.3mm}\frac{d\tPsigz}{ds}\hspace{0.3mm}s^3ds},   
               \\[12pt]
               \hspace{0.05cm}\displaystyle{-\!\int_0^{\infty}\!
               \bigg(
\hspace{0.2mm}\tilde\rho_{*1}+\frac{2}{3}\hspace{0.2mm}s^2\tilde\rho_{*2}
               \bigg)\hspace{-0.1mm}\frac{d\tPsigz}{ds}\hspace{0.3mm}s^3ds}, 
               \\[12pt]
               \hspace{0.05cm}\displaystyle{-\!\int_0^{\infty}\!\tilde\rho_{*0}\hspace{0.2mm}\frac{d}{ds}
               \bigg(
               \tPsigu+\frac{2}{3}\hspace{0.2mm}s^2\hspace{0.2mm}\tPsigd
               \bigg)\hspace{-0.1mm}s^3ds}, 
        \end{cases}
  \label{eq:w012_integrals}
  \end{equation}
  for $i=0,1,2$, from top to bottom, respectively.

  As well known, in multi-component systems the virial energy $W_*$ is 
  {\it not} the gravitational energy $U_*$ of the stellar component in 
  the total potential. In analogy with the previous discussion, we now 
  express explicitly the different contributions to the potential energy 
  $\Us$ of the stellar component. We write
  \begin{equation}
        \Us=\Ug+\Ubh=\Uss+\Udm+\Ubh.
  \end{equation}
  In particular,
  \begin{equation}
        \Ug=-\,\frac{1}{2}\int\!\rhos \Psis d^3\xv\,-\!\hspace{0.2mm}\int\!\rhos \Psidm d^3\xv 
           =\Bg-\hspace{0.25mm}\Uss,
  \end{equation}
  where
  \begin{equation}
        \Bg = -\!\hspace{0.2mm}\int\!\rhos \Psig d^3\xv.
  \end{equation} 
  $\Bg$ is useful in the theory of galactic flows. Indeed, 
  $L_{\rm grav} \propto |\Bg|$, where $L_{\rm grav}$ is the energy per 
  unit time to be given to the ISM (via, e.g., supernova explosions, or 
  thermalization of the velocity of stellar winds, or AGN feedback) in 
  order to steadily extract the ISM mass injected over the galaxy body 
  in the unit time (e.g., from evolving stars; see Pellegrini 2011, 
  Posacki et al. 2013).
  
  \begin{figure*}
  \includegraphics[width=0.493\linewidth]{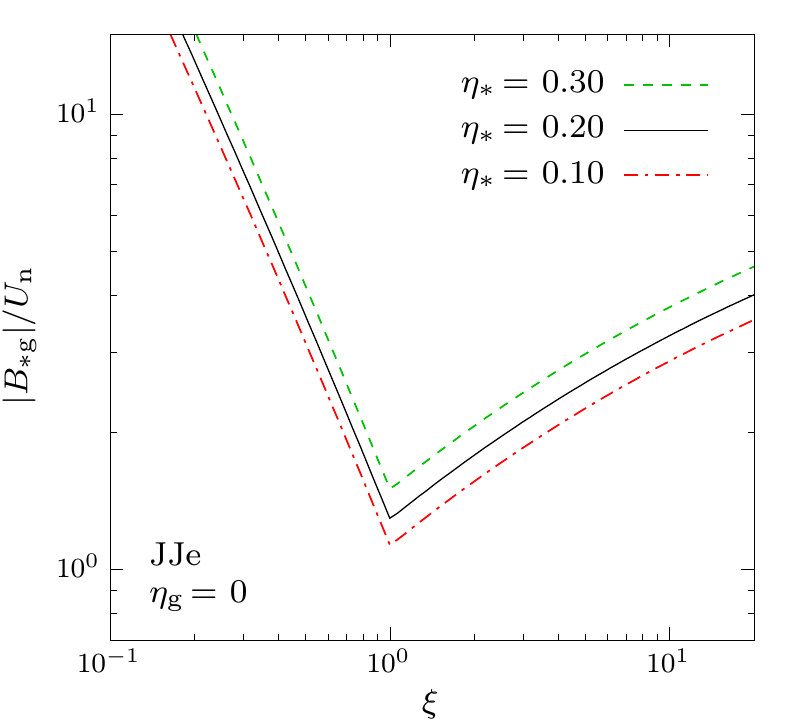}
  \hspace{0.8mm}
  \includegraphics[width=0.493\linewidth]{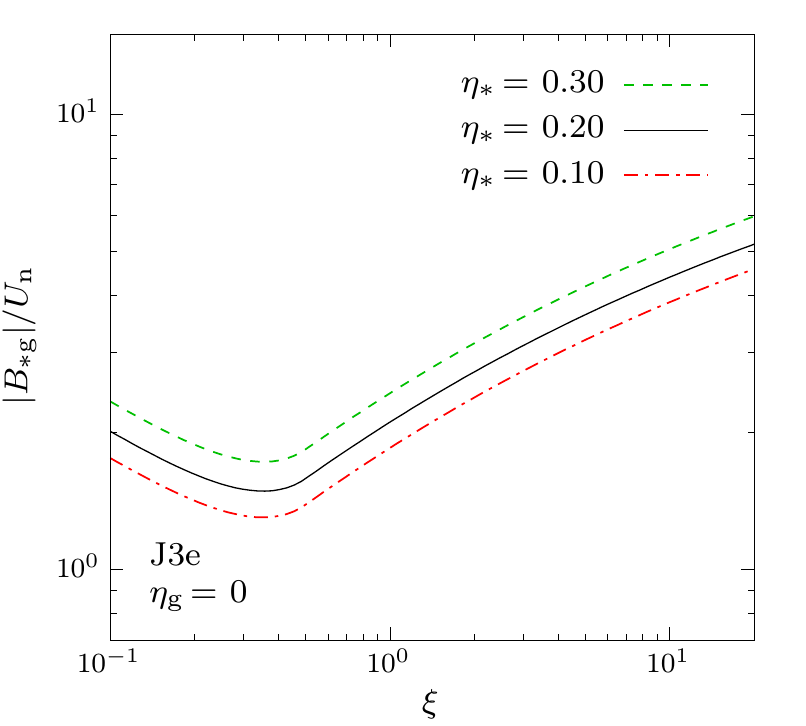}
  \vspace{-3mm}
  \caption{Absolute value of the energy $\Bg$, normalized to 
           $U_{\rm n} \equiv \Psin\Ms=G\Ms^2/\rs$, as a function of $\xi$, for three 
           representative minimum halo models, for a spherical galaxy without a central BH 
           (i.e., $\mu=0$). Left: JJe models; right: J3e models. Notice how for large values 
           of $\xi$ the curves in the two panels have similar behaviour. It can be proved 
           that for $\xi \to \infty$ they become identical (see Section \protect\ref{sec:VT}).}
  \label{fig:virial}
  \end{figure*}

  By using the homoeoidal expansion for $\rhos$ and $\Psig$, and changing 
  variables to spherical coordinates, an integration over the solid angle 
  shows that, limiting to linear terms in the flattenings,
  \begin{equation}
        \Bg = -\,\Un\MR\times\left(u_0 +\etas u_1 +\etag u_2\right)\!,
  \end{equation}
  where 
  \begin{equation}
         u_i =
         \begin{cases} 
               \hspace{0.05cm}\displaystyle{\int_0^{\infty}\!\tilde\rho_{*0}\tPsigz\,s^2ds},   
               \\[12pt]
               \hspace{0.05cm}\displaystyle{\int_0^{\infty}
              \bigg(
              \hspace{0.2mm}\tilde\rho_{*1}+\frac{2}{3}\hspace{0.2mm}s^2\tilde\rho_{*2}
              \bigg)\hspace{-0.1mm}\tPsigz\,s^2ds}, 
              \\[12pt]
              \hspace{0.05cm}\displaystyle{\int_0^{\infty}\!\tilde\rho_{*0} 
              \bigg(
              \tPsigu+\frac{2}{3}\hspace{0.2mm}s^2\hspace{0.1mm}\tPsigd
              \bigg)\hspace{-0.1mm}s^2ds}, 
         \end{cases}
  \label{eq:u012}
  \end{equation}
  for $i=0,1,2$, from top to bottom, respectively.  

  \subsection{The Virial Theorem for JJe models}

  For the JJe models the quantities $w_i$ in equation 
  \eqref{eq:w012_integrals} can be easily computed as
  \begin{equation}
         w_i(\JJe)=
         \begin{cases} 
               \hspace{0.05cm}\displaystyle{\frac{\xi-1-\ln\xi}{(\xi -1)^2}}, 
               \\[12pt]
               \hspace{0.05cm}\displaystyle{\frac{2(1- \xi) +(\xi+1)\hspace{-0.25mm}\ln\xi}{3(\xi -1)^3}}, 
               \\[12pt]
               \hspace{0.05cm}\displaystyle{\frac{\xi^2-1 -2\hspace{0.1mm}\xi\ln\xi}{3(\xi -1)^3}},
         \end{cases}
  \end{equation}
  and, for $\xi=1$,
  \begin{equation}
         w_0=\frac{1}{2}, \qquad\quad
         w_1=\frac{1}{18}, \qquad\quad 
         w_2=\frac{1}{9}.
  \end{equation}

  A few comments are in order. First, in case of $\etas=\etag=0$ the 
  function $\Wg$ reduces to that of spherical models already given in 
  CZ18. Second, for $\xi \to \infty$ the contribution to $\Wg$ in 
  equation \eqref{eq:w012} due to the function $w_1$ is subdominant 
  with respect those of $w_0$ and $w_2$, and we get 
  \begin{equation}
        \Wg \sim -\,\Un\MR\,\frac{3+\etag}{3\xi}.
  \end{equation}
  In the minimum halo case, form equation \eqref{eq:pos_cond_JJe} it 
  follows that $\Wg$ tends to a finite value depending only on the two 
  flattenings $\etas$ and $\etag$. The third comment concerns the 
  contribution of the DM to $\Wg =\Wss +\Wdm$ (where $\Wss$ is due to 
  the self-interaction of the stellar distribution, and $\Wdm$ to the 
  effect of the DM halo). From equation \eqref{eq:w012} we have the 
  estimate of $\Wg$ for small flattenings of the total and stellar 
  distributions. As the total density in JJe models is a Jaffe 
  ellipsoidal model as the stellar one, it is obvious that equation 
  \eqref{eq:w012} can be used {\it also} to estimate $\Wss$ for small 
  flattenings, just by considering in it $\MR=1$, $\xi=1$, and
  $\etag=\etas$, so that $\Wdm =\Wg -\Wss$.

  For what concerns the function $\Bg$, from equation \eqref{eq:u012} we 
  find
  \begin{equation}
         u_i(\JJe)=
         \begin{cases} 
               \hspace{0.05cm}\displaystyle{\frac{\ln\xi}{\xi -1}},  
               \\[12pt]
               \hspace{0.05cm}\displaystyle{\frac{\xi - 1 -\ln\xi}{3(\xi -1)^2}}, 
               \\[12pt]
               \hspace{0.05cm}\displaystyle{\frac{1- \xi +\xi\ln\xi}{3(\xi -1)^2}},
         \end{cases}
  \label{eq:u012_J3e}
  \end{equation}
  where, for $\xi=1$,
  \begin{equation}
         u_0=1, \qquad\quad 
         u_1=\frac{1}{6}, \qquad\quad 
         u_2=\frac{1}{6}.
  \end{equation}
  As for $\Wg$, in the limit case of $\etas=\etag=0$ also the function 
  $\Bg$ coincides with that found in CZ18. For large values of $\xi$ 
  (i.e., $\xi \to \infty$), also in case of $\Bg$ the spherical term and 
  that due to the total distribution dominate over the stellar one; the 
  resulting behaviour reads
  \begin{equation}
        \Bg \sim -\,\Un\MR\,\frac{3+\etag}{3\xi}\ln\xi
            \sim \Wg\ln\xi.
  \end{equation}
  Therefore, at variance with $\Wg$, the quantity $\Bg$ diverges for a DM 
  halo much more extended than the stellar distribution. Note that $\Uss$ 
  can also be obtained, for JJe models, not only from $\Wg$, but also from 
  $\Bg$, evaluating $\Bg/2$ with $\MR=1$, $\xi=1$, and $\etag=\etas$. Once 
  $\Uss$ is obtained, the evaluation of $\Udm$ through $\Bg$ is trivial, 
  since $\Udm=\Bg-2\Uss$. Fig. \ref{fig:virial} shows the trend of $|\Bg|$ 
  for a spherical galaxy, as a function of $\xi$, for different stellar 
  flattenings.

  \subsection{The Virial Theorem for J3e models}

  The quantities $w_0$, $w_1$ and $w_2$ are given by
  \begin{equation}
         w_i(\Jte)=
         \begin{cases} 
               \hspace{0.05cm}\displaystyle{\Hcsi(\xi,0)-\frac{\ln\xi}{\xi-1}}, 
               \\[11pt]
               \hspace{0.05cm}\displaystyle{\frac{1}{3}\hspace{0.3mm}\Hcsi(\xi,0)-\frac{\xi-1+(\xi-2)\ln\xi}{3(\xi-1)^2}}, 
               \\[11pt]
               \hspace{0.05cm}\displaystyle{\frac{\xi-1-\ln\xi}{3(\xi-1)^2}},
         \end{cases}
  \end{equation}
  where the function $\Hcsi(\xi,s)$ is defined as
  \begin{equation}
        \Hcsi(\xi,s) \equiv \int_s^{\infty}\ln\!\left(1+\frac{1}{t}\right)\!\frac{dt}{\xi+t},
  \end{equation}
  and it can be expressed in terms of the dilogarithm function (for more 
  details, see equation C1 in CMP19). In particular, for $\xi=1$,
  \begin{equation}
         w_0=\frac{\upi^2}{6}-1, \qquad 
         w_1=\frac{\upi^2}{18}-\frac{1}{2}, \qquad 
         w_2=\frac{1}{6}.
  \end{equation}
  In analogy with equation \eqref{eq:u012_J3e}, for J3e models the function 
  $\Bg$ is determined by
  \begin{equation}
         u_i(\Jte)=
         \begin{cases} 
               \hspace{0.05cm}\displaystyle{\Hcsi(\xi,0)}, 
               \\[9pt]
               \hspace{0.05cm}\displaystyle{\frac{1}{3}\hspace{0.3mm}\Hcsi(\xi,0)-\frac{\ln\xi}{3(\xi-1)}}, 
               \\[9pt]
               \hspace{0.05cm}\displaystyle{\frac{\ln\xi}{3(\xi-1)}},
         \end{cases}
  \end{equation}
  where, for $\xi=1$,
  \begin{equation}
         u_0=\frac{\upi^2}{6}, \qquad 
         u_1=\frac{\upi^2}{18}-\frac{1}{3}, \qquad 
         u_2=\frac{1}{3}.
  \end{equation}
  For $\etas=\etag=0$ the quantities $\Wg$ and $\Bg$ coincide with that 
  found in CMP19. Moreover, when considering large values of $\xi$, a 
  simple expansion shows that the asymptotic behaviours of $\Wg$ and $\Bg$ 
  are {\it identical} to that derived in case of JJe models (see Fig. 
  \ref{fig:virial}): a qualitative explanation is that in both cases the 
  Jaffe stellar distribution, for $\xi \to \infty$, is embedded in what we 
  call ``singular isothermal ellipsoid''.

  \section{Discussion and conclusions}\label{sec:Conclusions}

  In this paper we present two new families of two-component axially 
  symmetrical galaxy models: the ellipsoidal generalization of the 
  spherical JJ and J3 models, introduced in CZ18 and CMP19, respectively. 
  In both these new families the stellar density follows an ellipsoidal 
  Jaffe profile; then, in the JJe models the total density is described 
  by another ellipsoidal Jaffe law, in the J3e models the total density 
  is such that its difference with the stellar density (e.g. the 
  resulting DM halo) can be made similar to an ellipsoidal NFW model. 
  In both families the total density has a different flattening and scale 
  length with respect to the stellar density. Finally, a BH is also added 
  at the center. The JJe and J3e models are fully determined once the 
  stellar mass ($\Ms$) and the scale length of the stellar density 
  ($\rs$) are assigned, together with the total-to-stellar scale length 
  ratio ($\xi$), the total-to-stellar density ratio ($\MR$), the 
  flattening of the stellar profile ($\etas$), the flattening of the 
  total profile ($\etag$), and finally a BH-to-stellar mass ratio ($\mu$).
 
  One of the main advantages of the JJe and J3e models is that, thanks to 
  a homoeoidal expansion to the first order adopted in this paper, an 
  analytical treatment of several quantities of interest in theoretical and 
  observational works is possible, as detailed below.
  \begin{enumerate}
        \item The constraints on $\MR$ and $\xi$ to assure the positivity of the DM halo 
              density profile are derived analytically. For a given $\xi$, the model with the 
              minimum value $\Rm$ allowed for $\MR$ is called {\it minimum halo model}. A 
              general method to discuss the positivity of the DM distribution, defined as the 
              difference between two arbitrary ellipsoidal distributions, is also presented
              in the Appendix.
        \item We expand the density and potential profiles for a small deviation from spherical 
              symmetry by adopting the so-called {\it homoeoidal expansion method} (see CB05). 
              The derived analytical expressions correspond to positive densities if 
              $\etas \leq 1/3$, and $\etag \leq 1/3$ (JJe) or $\etag \leq 1/2$ (J3e). 
              The circular velocity $\vcirc$ in the equatorial plane ($z=0$) is also obtained.
        \item Using the homoeoidal expansions, we analytically solve the two-integral Jeans 
              equations, where the Satoh (1980) $k$-decomposition is adopted  to split the 
              azimuthal velocity field in its ordered ($\vphib$) and random ($\sigphi$) components. 
              The solutions are given at their first order terms in the flattening; the asymptotic 
              expansion near the center and in the outer regions are presented, and maps of $\sigs$, 
              $\sigphi$ and $\vphib$ in the meridional plane for a representative galaxy are also 
              shown. The $\sigs$ contour lines are elongated along the $z$-axis. For fixed 
              distance from the galactic center, the values of $\sigs$, $\sigphi$ and $\vphib$ 
              keep larger for the J3e model than for the JJe one, due to the broader DM 
              distribution of the former models. Finally, the link between the angular momentum 
              $J_z(R,z)$ away from the equatorial plane, and that on the plane 
              $\Jc(R\;\!)$, is 
              briefly presented; this link is useful to consider for problems involving infalling 
              gas that conserves its angular momentum.
        \item Finally, the analytical expressions for the quantities entering the Virial Theorem, 
              such as the interaction energy and the potential energies, are derived as a function 
              of the model parameters. 
  \end{enumerate}

  The JJe and J3e models represent a significant improvement over the 
  spherical counterparts discussed in CZ18 and CMP19: they provide a more 
  advanced modelling of the dynamics of elliptical galaxies, while still 
  keeping realistic  mass distributions, as are the Jaffe or the NFW laws. 
  The analytical formulation of all their dynamical properties (e.g. 
  kinematical quantities and virial energies) can be used to understand what 
  is the effect of the various parameters (flattening, scale lengths, mass 
  ratio, rotational support) in determining these properties. The analytical 
  expressions are also useful when realistic axially symmetric two-component 
  galaxy models are needed, for example to be given in input in numerical 
  simulations, as those reproducing galactic flows (see e.g. Gan et al. 2019).

  \section*{Acknowledgements}

  We thanks Caterina Caravita, Zhaoming Gan, and Federico Marinacci for careful 
  and independent numerical checks of several formulae. The anonymous referee
  is warmly thanked for several suggestions that considerably improved the paper.
  
  \section*{Data Availability}

  No datasets were generated or analysed in support of this research.



  \appendix
  \section{Positivity of the DM distribution}\label{app:pos}

  In order to discuss the positivity of the DM distribution of JJe 
  and J3e models, it is convenient to set up the problem in the more 
  general case of two arbitrary ellipsoidal distributions, and then 
  to specialized the results to the specific cases. Let 
  \begin{equation}
        \rhos=\Es(\ms), 
        \qquad\quad 
        \rhog=\MR\hspace{0.2mm}\Eg(\mg),
  \label{eq:Es_Eg}
  \end{equation}
  be the stellar and total density distributions, with $\Es$ and $\Eg$ 
  the arbitrary functions describing the profiles, where $\ms$ and $\mg$ 
  are defined in Section \ref{sec:Models}. We change variables from
  $(R,z)$ to $(r\sin\theta,r\cos\theta)$, so that
  \begin{equation}
        \ms=s\,\Os, \qquad\,\, 
        \mg=s\,\Og, \qquad\,\, 
        s=\frac{r}{\rs},
  \end{equation}
  and 
  \begin{equation}
        \Os^2=\sin^2\!\theta+\frac{\cos^2\!\theta}{\qs^2}, 
        \qquad\,\,
        \Og^2=\sin^2\!\theta+\frac{\cos^2\!\theta}{\qg^2}.
  \label{Os_Og}
  \end{equation}

  The positivity condition for $\rhoDM = \rhog-\rhos$ becomes
  \begin{equation}
        \MR \geq \Rm = \sup_{\I}\Hf(s,\theta), \qquad \Hf(s,\theta) 
                     = \frac{\Es(\ms)}{\Eg(\mg)},
  \label{eq:pos_cond}
  \end{equation}
  where, from Fig. \ref{fig:pos}, 
  $\I \equiv \left\lbrace (s,\theta)\,|\, s \geq 0,\, 0\leq\theta\leq\upi/2 \right\rbrace$: 
  we restrict to values of $\theta$ between $0$ and $\upi/2$ since 
  $\Hf(s,\upi-\theta)=\Hf(s,\theta)$. A DM halo with $\MR=\Rm$ is called 
  a {\it minumim halo}: clearly, if $\MR$ decreases slightly below $\Rm$, the 
  DM density becomes first negative at the position where $\Hf(s,\theta)=\Rm$. 
  It follows that
  \begin{equation}
        \Rm=\max\hspace{0.5mm}(\MRc,\MRinf,\MR_0,\MReq,\MRint),
  \label{eq:Rm_max(...)}
  \end{equation}
  where the values in parentheses are the $\sup\Hf(s,\theta)$ over the 
  corresponding regions in Fig. \ref{fig:pos}. Geometrically, 
  $\sup\Hf(s,\theta)$ can be located only at the center, at infinity, on the 
  equatorial plane, along the symmetry axis, or in the interior.  

  We first show that for $\qs\neq\qg$ the function $\Hf$ has no critical 
  points in $\intI$. Indeed, a simple computation shows that its gradient can 
  vanish when
  \begin{equation}
         \begin{cases}
               \hspace{0.05cm}\displaystyle{\Os\hspace{0.1mm}\frac{d\Es}{d\ms}\hspace{0.1mm}\hspace{0.1mm}\Eg=
\Og\hspace{0.1mm}\frac{d\Eg}{d\mg}\hspace{0.1mm}\hspace{0.1mm}\Es}, 
               \\[15pt]
               \hspace{0.05cm}\displaystyle{\frac{d\Os}{d\theta}\frac{d\Es}{d\ms}\hspace{0.1mm}\Eg=
\frac{d\Og}{d\theta}\frac{d\Eg}{d\mg}\hspace{0.1mm}\Es},
         \end{cases}
  \label{eq:dH/ds_dH/dtheta}
  \end{equation}
  where the first equation corresponds to $\partial\Hf/\partial s=0$, and 
  the second to $\partial\Hf/\partial \theta=0$. The proof proceeds as 
  follows. If the first equation is not satisfied in $\intI$, there is 
  nothing to prove. So, let us assume that the first identity is satisfied 
  somewhere in $\intI$. Then, for non-negative and monotonically decreasing 
  density distributions, the second equation reduces to
  \begin{equation}
        \frac{d\Os}{d\theta}\,\Og=\Os\hspace{0.1mm}\frac{d\Og}{d\theta};
  \end{equation}
  however, it is trivial to show that, for $\qs \neq \qg$, there are no 
  solutions for $0<\theta<\upi/2$. We are left with the case $\qs=\qg$. In 
  this circumstance, the two equations of the system \eqref{eq:dH/ds_dH/dtheta} 
  become coincident. As a consequence, $\MRint$ must be determined by solving 
  the equation $\partial\Hf/\partial s=0$, and imposing the condition 
  $\qs\!=\qg$. In particular, we note that in the special case $\qs=\qg$ the 
  problem formally reduces to the study of the positivity in spherical systems 
  (see e.g. CZ18; CMP19).

  \subsection{The positivity condition for JJe models}

  \begin{figure}
  \centering
  \begin{tikzpicture}
        \draw[thick] (0,1.57) node (yaxis) [left] {\Large $\frac{\upi}{2}$};
        \draw[thick] (0,0) node (yaxis) [left] {\normalsize $0$};
        \draw[->] (0,0) -- (7.1,0) node[right] {\large $s$};
        \draw[->] (0,0) -- (0,2.4) node[above] {\large $\theta$};
        \draw[samples=400,scale=1,domain=0:6.2,smooth,variable=\x,blue,line width=0.25mm] plot ({\x}, 1.57);
        \draw[samples=400,scale=1,domain=6.2:7,dashed,variable=\x,blue,line width=0.25mm] plot ({\x}, 1.57);
        \draw[samples=400,scale=1,domain=0:6.2,smooth,variable=\x,blue,line width=0.25mm] plot ({\x}, 0);
        \draw[samples=400,scale=1,domain=6.2:7,dashed,variable=\x,blue,line width=0.25mm] plot ({\x}, 0);
        \draw[blue, line width=0.25mm] (0,0) -- (0,1.57);
        \fill[gray!40,nearly transparent] (0,0) -- (0,1.57) -- (6.95,1.57) -- (6.95,0) -- cycle;
        \node[text width=0cm] at (-0.45,0.785) {\large $\Ic$};
        \node[text width=0cm] at (2.2,1.88) {\large $\Ieq\,({\rm equatorial \; plane})$};
        \node[text width=0cm] at (7.1,0.785) {\large $\Iinf$};
        \node[text width=0cm] at (2.79,-0.325) {\large $\Izero\;(z\,{\rm -\,axis})$};
        \node[text width=0cm] at (3.27,0.785) {\large $\intI$};
  \end{tikzpicture}
  \vspace{-2mm}
  \caption{Illustration of the region $\I$ over which the function $\Hf(s,\theta)$ must be maximized 
           in order to guarantee positivity of the DM density distribution $\rhoDM=\MR\Eg(\mg)-\Es(\ms)$.}
  \label{fig:pos}
  \end{figure}

  We now apply the previous considerations to the ellipsoidal generalization 
  of the spherical two-component $\gamma$ models in CZ18, where 
  \begin{equation}
        \Hf(s,\theta)=\frac{1}{\xi\alpha}\!\left(\frac{\Og}{\Os}\hspace{-0.4mm}\right)^{\hspace{-1mm}\gamma}\!\!\left(\frac{\xi+s\,\Og}{1+s\,\Os}\right)^{\hspace{-1mm}4-\gamma}\!, 
                     \quad\,\, 
                     \alpha \equiv \frac{\qs}{\qg},
  \label{eq:F_JJe}
  \end{equation}
  and $0 \leq \gamma <3$.

  We start with the discussion of the positivity condition on the 
  boundary of $\I$ (see Fig. \ref{fig:pos}). Along $\Ic$, 
  \begin{equation}
        \MRc = \frac{\xi^{3-\gamma}}{\alpha}\times\max_{0\leq\theta\leq\frac{\upi}{2}}\!f^{\hspace{0.1mm}\gamma}\hspace{-0.2mm}(\theta),
            \qquad
            f(\theta) \equiv \frac{\Og}{\Os}\hspace{-0.4mm}, 
  \label{eq:Rc_gammagamma_e}
  \end{equation}
  so that the problem reduces to the study of 
  \begin{equation} 
        \frac{df}{d\theta} \propto 
        \big(
        \hspace{0.1mm}1-\alpha^2\hspace{0.1mm}
        \big)\!\hspace{0.2mm}\sin 2\theta.
  \end{equation}
  For $\qs<\qg$ (i.e. $\alpha<1$), the maximum ($1$) is reached at 
  $\theta=\upi/2$, while for $\qs>\qg$ (i.e. $\alpha>1$) the maximum 
  ($\alpha^{\gamma}$) is reached at $\theta=0$. Summarizing,
  \begin{equation} 
        \MRc=\xi^{3-\gamma}\times
        \max\!\hspace{0.3mm}\left(\frac{1}{\alpha},\,\alpha^{\gamma-1}\right)\!.
  \label{eq:pos_Ic_gammagamma_e}
  \end{equation} 
  Over $\Iinf$ (i.e., for $s \to \infty$), from a similar analysis, 
  \begin{equation}
        \MRinf=\frac{1}{\xi\alpha}\times\max_{0\leq\theta\leq\frac{\upi}{2}}\!f^{\hspace{0.15mm}4}\hspace{-0.2mm}(\theta)
              =\,\frac{1}{\xi}\times\max\!\hspace{0.3mm}\left(\frac{1}{\alpha},\,\alpha^3\right)\!.
  \end{equation}
  The positivity along the symmetry axis $\Izero$, and on the equatorial 
  plane $\Ieq$, requires 
  \begin{equation}
        \MR_0=\frac{\alpha^3}{\xi}\sup_{s \geq 0}\!\hspace{0.3mm}\left(\frac{\xi\qg+s}{\qs+s}\right)^{\!\!4-\gamma}\!
             =\,\max\!\hspace{0.3mm}\left(\frac{\alpha^3}{\xi},\,\xi^{3-\gamma}\alpha^{\gamma-1}\right)\!,
  \label{eq:R0_gammagamma_e}
  \end{equation}
  and
  \begin{equation}
        \MReq=\frac{1}{\xi\alpha}\sup_{s \geq 0}\!\hspace{0.3mm}\left(\frac{\xi+s}{1+s}\right)^{\!\!4-\gamma}\!
             =\frac{1}{\alpha}\times\max\!\hspace{0.3mm}\left(\frac{1}{\xi},\,\xi^{3-\gamma}\right)\!.
  \label{eq:Req_gammagamma_e}
  \end{equation}
  Finally, we consider $\intI$, and, according to equation 
  \eqref{eq:dH/ds_dH/dtheta}, only for $\qs=\qg$ (i.e., $\alpha=1$). Under 
  this condition, the study of equation \eqref{eq:F_JJe} is trivial, and it 
  shows that no maxima are contained in $\intI$, even in this case. The 
  positivity condition in the special case $\alpha=1$ is then obtained from 
  equation \eqref{eq:Rm_max(...)}, and it reads 
  \begin{equation}
        \MR\geq\Rm=\max\!\hspace{0.3mm}\left(\frac{1}{\xi},\,\xi^{3-\gamma}\right)\!,
        \qquad\,\,\, 
        (\qs=\qg),
  \end{equation}
  in agreement with the result for spherical JJ models in CZ18.

  \subsection{The positivity condition for J3e models}

  Equation \eqref{eq:F_JJe} becomes
  \begin{equation}
        \Hf(s,\theta)=\frac{1}{\alpha}\!\left(\frac{\Og}{\Os}\hspace{-0.4mm}\right)^{\hspace{-1mm}2}\!\frac{\xi\hspace{0.15mm}+s\,\Og}{(1+s\,\Os)^2}, 
        \quad\,\, 
        \alpha \equiv \frac{\qs}{\qg}.
  \end{equation}

  Repeating the same treatment of JJe models, we immediately obtain  
  \begin{equation}
        \MRc=\frac{\xi}{\alpha}\times\max_{0\leq\theta\leq\frac{\upi}{2}}\!\left(\frac{\Og}{\Os}\hspace{-0.4mm}\right)^{\hspace{-1mm}2}\!
            =\,\xi\times\max\!\hspace{0.3mm}\left(\frac{1}{\alpha},\,\alpha\right)\!;
  \label{eq:pos_Ic_J3_e}
  \end{equation}
  moreover, as in J3e models the total density profile decreases more slowly 
  than the stellar density for $s \to \infty$, positivity at large radii is 
  assured independently on the value of $\MR$, so that formally $\MRinf=0$. 
  Along the symmetry axis $\Izero$, 
  \begin{equation}
        \MR_0=\qg\alpha^3\sup_{s \geq 0}\hspace{0.3mm}\frac{\xi\qg+s}{(\qs+s)^2}
             =\alpha\times
              \begin{cases} 
                    \hspace{0.05cm}\displaystyle{\frac{\alpha^2}{4(\alpha-\xi)}}, & \hspace{0.1cm} \displaystyle \xi \leq \frac{\alpha}{2}, 
                    \\[8pt]
                    \hspace{0.05cm}\displaystyle{\xi}, & \hspace{0.1cm} \xi \geq \displaystyle \frac{\alpha}{2},
              \end{cases}
  \end{equation}
  and, along the equatorial plane $\Ieq$,
  \begin{equation}
        \MReq=\frac{1}{\alpha}\,\sup_{s \geq 0}\hspace{0.3mm}\frac{\xi+s}{(1+s)^2}
             =\frac{1}{\alpha}\times
              \begin{cases} 
                    \hspace{0.05cm}\displaystyle{\frac{1}{4(1-\xi)}}, & \hspace{0.1cm} \displaystyle \xi \leq \frac{1}{2}, 
                    \\[8pt]
                    \hspace{0.05cm}\displaystyle{\xi}, & \hspace{0.1cm} \xi \geq \displaystyle \frac{1}{2}.
              \end{cases}
  \end{equation}
  For what concerns the positivity in the interior of $\I$, the only case 
  to be considered is $\qs=\qg$ (i.e., $\alpha=1$). It is easy to show that 
  $\Hf$ has no critical points in $\intI$ when $\xi>1/2$; for $\xi \leq 1/2$, 
  instead, $\MRint=1/[4(1-\xi)]$. In conclusion, for $\alpha \neq 1$, $\MR$ 
  is obtained from equation \eqref{eq:Rm_max(...)} and the previous results,
  while in the special case $\alpha=1$ the final condition is
  \begin{equation}
        \MR\geq\Rm=
        \begin{cases} 
              \hspace{0.05cm}\displaystyle{\frac{1}{4(1-\xi)}}, & \hspace{0.3cm} \displaystyle \xi \leq \frac{1}{2}, 
              \\[8pt]
              \hspace{0.05cm}\displaystyle{\xi}, & \hspace{0.3cm} \xi \geq \displaystyle \frac{1}{2},
        \end{cases}
        \qquad\, 
        (\qs=\qg),
  \end{equation}
  in agreement with the result for spherical J3 models in CMP19.

  \section{Homoeoidal expansion}\label{app:Hom_Exp} 

  A thorough description of the homoeoidal expansion method can be 
  found in CB05. Here we just report the formulae strictly needed for 
  the present work. Consider an ellipsoidal mass density distribution 
  $\rho$ stratified, in Cartesian coordinates, over surfaces labelled by
  \begin{equation}
         m^2 \equiv \frac{x^2}{a^2}+\frac{y^2}{b^2}+\frac{z^2}{c^2}
                =   \frac{x^2}{a^2}+\frac{y^2}{a^2(1-\epsilon)^2}+\frac{z^2}{a^2(1-\eta)^2},
  \end{equation}
  where $a \geq b \geq c > 0$, $ b/a \equiv 1 -\epsilon $, and $c/a
  \equiv 1-\eta$; when $\epsilon=\eta=0$, $m=s=r/a$. We write  
  \begin{equation}
        \rho(m)=\rhon\!\times\frac{\rhotil(m)}{(1-\epsilon)(1-\eta)},
        \qquad 
        \rhon\equiv\frac{M_{\rm n}}{4\upi a^3},
  \label{eq:rho_Hom_exp}
  \end{equation}
  where $\rhon$ is a normalization density, and $M_{\rm n}$ is the mass 
  of the ellipsoid contained inside the arbitrary ellipsoid defined by 
  $m$. With this choice, $M_{\rm n}$ is independent of the adopted 
  flattenings, the so-called {\it constrained case}. Of course, for a 
  model of finite total mass, the natural choice is to adopt for $M_{\rm n}$ 
  the total mass. Note that, for a model of finite total mass $M$, the 
  normalization assures that the total mass is conserved independently of 
  the value of $\epsilon$ and $\eta$. In case of an infinite total mass 
  (such as $\rhog$ in J3e models), the condition assures that the mass 
  contained inside any $m$ is conserved. The oblate axisymmetric 
  models discussed in this paper (see Section \ref{sec:Hom_Exp}) are 
  obtained for $\epsilon =0$, $a=\rs$, $\eta=1-\qs$
  for the stellar component, and $\eta=1-\qg$ for the total
  density.

  By expanding at the linear order in terms of the flattenings one obtains  
  \begin{equation}
        \frac{\rhotil(m)}{(1-\epsilon)(1-\eta)} = 
        \vrtz(s)+(\epsilon +\eta)\!\hspace{0.5mm}\vrtu(s)+\big(\epsilon\tilde{y}^2+\,\eta \ztil^2\big)\!\hspace{0.4mm}\vrtd(s),
  \label{eq:vrhotil_012}
  \end{equation}
  where $\tilde y\equiv y/a$, $\tilde z\equiv z/a$, and
  \begin{equation}
        \vrtz(s)=\vrtu(s)=\rhotil(s), 
        \qquad\quad
        \vrtd(s)=\frac{1}{s}\frac{d\rhotil(s)}{ds}.
  \end{equation}

  In order to be physically acceptable, the expanded density must be
  nowhere negative, and this requirement sets an upper limit on the 
  possible values of $\epsilon$ and $\eta$, as a function of the specific 
  density profile adopted. By changing variables to spherical coordinates, 
  and following the approach introduced in Appendix \ref{app:pos}, it can 
  be shown that positivity of equation \eqref{eq:vrhotil_012} for 
  $0\leq\epsilon\leq\eta<1$, and for a monothonically decreasing 
  $\tilde\rho (s)$, is assured provided that
  \begin{equation}
        \epsilon \geq (A_M-1)\eta-1,
        \qquad
        A_M\equiv\sup_{s\geq 0}\left|\frac{d\ln\tilde\rho(s)}{d\ln s}\right|\!.
  \label{eq:pos_cond_general}
  \end{equation}
  For $\gamma$ models, $A_M= 4$, so that in the axysimmetric case
  ($\epsilon =0$) we recover the condition $\eta\leq 1/3$ (see CB05). 

  The general quadrature formula for the potential of a density 
  distribution $\rho (m)$ is given by 
  \begin{equation}
        \Psi(\xv)=\upi a\hspace{0.1mm}b\hspace{0.1mm}c\hspace{0.3mm}G \int_0^\infty 
\frac{\Delta\!\hspace{0.3mm}\Psi[m(\xv;\tau)]}{\sqrt{(a^2+\tau)(b^2+\tau)(c^2+\tau)}}\,d\tau,
  \label{eq:Chandra}
  \end{equation}
  (see e.g. Kellogg 1953; Chandrasekhar 1969; BT08), where
  \begin{equation}
        \Delta\!\hspace{0.3mm}\Psi[m(\xv;\tau)]\equiv2 \int_{m(\xv;\tau)}^\infty \rho(m)m dm,
  \end{equation}
  and
  \begin{equation}
         m^2(\xv;\tau) \equiv \frac{x^2}{a^2+\tau}+\frac{y^2}{b^2+\tau}+\frac{z^2}{c^2+\tau};
  \end{equation}
  note that the variable $\tau$ has the dimension of a squared length.

  By inserting equation \eqref{eq:rho_Hom_exp} in equation 
  \eqref{eq:Chandra}, and after normalization of all lengths to $a$ (and 
  $\tau$ to $a^2$), it is immediate to show that 
  \begin{equation}
        \Psi(\xv)=\Psin\!\times\tilde\Psi(\xv), 
        \qquad 
        \Psin\equiv\frac{GM_{\rm n}}{a},
  \end{equation}
  where the meaning of the function $\tilde\Psi$ is obvious. Expanding
  the integrand in equation \eqref{eq:Chandra} at linear order in the 
  flattenings, and inverting order of integration, some algebra shows that   
  \begin{equation}
        \tilde{\Psi}(\xv)=\tpsiz (s)+(\epsilon +\eta)\hspace{0.2mm}\tpsiu
(s)+\big(\epsilon {\tilde y}^2+ \eta \ztil^2\big)\tpsid (s),
  \label{eq:Psitilde_expansion}
  \end{equation}
  where
  \begin{equation}
        \tpsii(s)=
        \begin{dcases}
               \frac{1}{s} \int_0^s\rhotil(m)m^2dm \,+ \int_s^{\infty}
\rhotil(m)mdm, \\
\frac{1}{3s^3} \int_0^s\rhotil(m)m^4dm + \frac{1}{3}\int_s^{\infty} \rhotil(m)mdm, \\
-\,\frac{1}{s^5} \int_0^s \rhotil(m)m^4dm ,\\
        \end{dcases}
  \label{eq:Psi_i_012}
  \end{equation}
  with $i=0,1,2$, respectively. Then, Poisson's equation for the 
  dimensionless potential-density pair $(\tilde{\Psi},\rhotil)$ becomes
  \begin{equation}
        \lapn\tilde{\Psi}(\xv)=-\,\frac{\rhotil(m)}{(1-\epsilon)(1-\eta)}, 
        \qquad
        \lapn \equiv a^2\nabla^2.
  \label{eq:Poisson}
  \end{equation}
  The previous formulae, in the axisymmetric oblate case, are obtaines 
  by setting $\epsilon=0$ and $0<\eta<1$. It may be convenient in some 
  computation to recast 
  equations \eqref{eq:vrhotil_012} and \eqref{eq:Psitilde_expansion} in 
  terms of $R^2$ instead of $z^2=r^2-R^2$, and in this case the 
  corresponding functions are given by 
  \begin{equation}
        \rtz=\vrtz, \qquad\, 
        \rtu=\vrtu+s^2\vrtd, \qquad\, 
        \rtd=-\,\vrtd,
  \end{equation}
  and
  \begin{equation}
        \tPsiz=\tpsiz, \qquad\, 
        \tPsiu=\tpsiu+s^2\tpsid, \qquad\, 
        \tPsid=-\,\tpsid.
  \end{equation}
  For example, when computing properties on the equatorial plane, 
  where $z=0$, such as for example in the derivation of 
  the circular speed $\vcirc(R\hspace{0.25mm})$, angular momentum 
  $\Jc(R\hspace{0.25mm})$, or radial epicyclic frequency 
  $\krad(R\hspace{0.25mm})$, it is useful to work with the 
  ``explicit\hspace{0.2mm}-$z$ formulation'', while in some other case, 
  such as the integration of the Jeans equations, or the derivation of 
  the vertical epicyclic frequency $\kvert(R\hspace{0.25mm})$, it is more 
  useful to use the 
  ``explicit\hspace{0.2mm}-$R$ formulation''.
  In particular, we recall that $\krad$ and $\kvert$ are defined as
  \begin{equation}
        \krad^2(R\hspace{0.25mm})\equiv\frac{1}{R^3}\frac{d\Jc^2}{dR},
        \qquad
        \kvert^2(R\hspace{0.25mm})\equiv-\hspace{0.35mm}\bigg(\frac{\partial^2\Psi}{\partial z^2}\bigg)_{\hspace{-0.8mm}z=0}.
  \end{equation}
  By defining $\kn^2\equiv\Psin/a^2$, simple algebra shows that
  \begin{equation}
        \frac{\kvert^2(R\hspace{0.25mm})}{\kn^2}
=\kverttz^2(\Rtil\hspace{0.25mm})+\eta\hspace{0.35mm}\kverttu^2(\Rtil\hspace{0.25mm})+\eta\hspace{0.15mm}\Rtil^2\kverttd^2(\Rtil\hspace{0.25mm}),
  \end{equation}
  where
  \begin{equation}
        \kvertti^2(\Rtil\hspace{0.25mm}) \equiv -\hspace{0.2mm}\frac{1}{\Rtil}\frac{d\tilde{\Psi}_i(\Rtil\hspace{0.25mm})}{d\Rtil},
        \qquad\,\, (i=0,1,2),
  \end{equation}
  and
  \begin{equation}
        \frac{\krad^2(R\hspace{0.25mm})}{\kn^2}
=\kradtz^2(\Rtil\hspace{0.25mm})+\eta\hspace{0.35mm}\kradtu^2(\Rtil\hspace{0.25mm}),
\end{equation}
  with
  \begin{equation}
        \kradti^2(\Rtil\hspace{0.25mm}) \equiv -\hspace{0.3mm}\frac{1}{\Rtil^3}\frac{d}{d\Rtil}\!\left[\Rtil^3\hspace{0.2mm}\frac{d\tilde{\psi}_i(\Rtil\hspace{0.25mm})}{d\Rtil}\right]\!,
        \qquad\, (i=0,1).
  \end{equation}

  \onecolumn
  \section{Velocity dispersion}\label{app:ABCDEFGH}

  We report here the explicit expressions of the functions entering the
  velocity dispersion profiles in Section \ref{sec:Solution}. 

\subsection{JJe models}
The three functions
describing the BH contribution in equation \eqref{eq:ABC} are
\begin{equation}
A(s)\equiv
\int_s^{\infty}\frac{\rtsz}{s'^{\hspace{0.15mm}2}}\hspace{0.5mm}ds'\hspace{-0.15mm}=\hspace{0.1mm}
\frac{12s^3 + 6s^2 - 2s + 1}{3s^3(1+s)}+4\ln\frac{s}{1+s},
\end{equation}
\begin{equation}
B(s)\equiv
\int_s^{\infty}\frac{\rtsu}{s'^{\hspace{0.15mm}2}}\hspace{0.5mm}ds'\hspace{-0.15mm}=\hspace{0.1mm}
\frac{24 s^4 +36 s^3 + 8 s^2 - 2 s -1}{3s^3(1+s)^2}+8\ln\frac{s}{1+s},
\end{equation}
\begin{equation}
C(s)\equiv
\int_s^{\infty}\frac{\rtsd}{s'^{\hspace{0.15mm}2}}\hspace{0.5mm}ds'\hspace{-0.15mm}=\hspace{0.1mm}
-\,\frac{180 s^6 +270 s^5 + 60 s^4 -15 s^3 + 6 s^2 -3 s - 4}{10s^5(1+s)^2}-18\ln\frac{s}{1+s}.
\end{equation}
Note that from equation \eqref{eq:rtgi_JJe_J3e} it follows that $A>0$, $B<0$ and $C>0$.

For the contribution of the galaxy to the velocity dispersion in equation \eqref{eq:DEFGH},
an elementary integration leads to
\begin{equation}
D(s)\equiv
-\int_s^{\infty}\hspace{-0.25mm}\rtsz\hspace{0.15mm}\frac{d\tPsigz}{ds'}\hspace{0.3mm}ds'=
-\,\frac{3\xi^2-\xi-1}{\xi^2(\xi-1)(1+s)}-\frac{(3\xi+2)s-\xi}{2\xi^2s^2(1+s)}
-\frac{1}{\xi^3(\xi-1)^2}\ln\frac{s}{\xi+s}-\frac{3\xi-4}{(\xi-1)^2}\ln\frac{s}{1+s},
\end{equation}
\begin{equation*}
E(s)\equiv
-\int_s^{\infty}\hspace{-0.25mm}\rtsu\hspace{0.15mm}\frac{d\tPsigz}{ds'}\hspace{0.3mm}ds'=
-\,\frac{2(3\xi^3-6\xi^2+2\xi-1)s + 9\xi^3-18\xi^2+9\xi-4}{2\xi^2(\xi-1)^2(1+s)^2}
-\frac{2(\xi-1)s+\xi}{2\xi^2s^2(1+s)^2}
\end{equation*}
\begin{equation}
+\,\frac{3\xi-1}{\xi^3(\xi-1)^3}\ln\frac{s}{\xi+s} 
-\frac{3\xi^2-9\xi+8}{(\xi-1)^3}\ln\frac{s}{1+s},
\end{equation}
\begin{equation*}
F(s)\equiv
-\int_s^{\infty}\hspace{-0.25mm}\rtsd\hspace{0.15mm}\frac{d\tPsigz}{ds'}\hspace{0.3mm}ds'=
\frac{2(5\xi^5-8\xi^4+\xi^3+\xi^2+\xi-1)s +15\xi^5-24\xi^4+3\xi^3+3\xi^2+5\xi-4}{\xi^4(\xi-1)^2(1+s)^2}
\end{equation*}
\begin{equation}
+\,\frac{4(5\xi^3+2\xi^2-3)s^3 -\xi(5\xi^2+2\xi-6)s^2 + 2\xi^2(\xi-2)s+3\xi^3}{6\xi^4s^4 (1+s)^2}-
\frac{2(2\xi-1)}{\xi^5(\xi-1)^3}\ln\frac{s}{\xi +s} +
\frac{2(5\xi^2-13\xi+9)}{(\xi-1)^3}\ln\frac{s}{1+s},
\end{equation}
\begin{equation*}
G(s)\equiv
-\int_s^{\infty}\hspace{-0.25mm}\rtsz\hspace{0.15mm}\frac{d\tPsigu}{ds'}\hspace{0.3mm}ds'=
\left[24\ln\frac{s}{1+s}+\frac{120s^5+60s^4-20s^3+10s^2-6s+4}{5s^5(1+s)}\right]\!\xi^2\ln\frac{\xi+s}{\xi}+\frac{5\xi^2+10\xi+9}{5\xi^3}\ln\frac{\xi+s}{s}
\end{equation*}
\begin{equation*}
+\,\frac{\xi(24\xi^3-60\xi^2+43\xi-5)}{(\xi-1)^3}\ln\frac{\xi+s}{1+s}-\frac{1}{\xi^2(\xi-1)^2(\xi+s)}+\frac{\xi(4\xi^2-6\xi+1)}{(\xi-1)^2(1+s)}
\end{equation*}
\begin{equation}
+\,\frac{4(20\xi^3+5\xi^2-2)s^3-\xi(40\xi^2+10\xi+1)s^2+4\xi^2(5\xi+1)s-8\xi^3}{10\xi^2s^4}-24\xi^2\Hcsi(\xi,s),
\end{equation}
\begin{equation*}
H(s)\equiv
-\int_s^{\infty}\hspace{-0.25mm}\rtsz\hspace{0.15mm}\frac{d\tPsigd}{ds'}\hspace{0.3mm}ds'=
\left[80\ln\frac{1+s}{s}-\frac{1680s^7+840s^6-280s^5+140s^4-84s^3+56s^2-40s+30}{21s^7(1+s)}\right]\!\xi^2\ln\frac{\xi+s}{\xi}
\end{equation*}
\begin{equation*}
+\,\frac{2(35\xi^4+21\xi^3-14\xi-15)}{21\xi^5}\ln\frac{\xi+s}{s}-\frac{2\xi(40\xi^3-100\xi^2+75\xi-14)}{(\xi-1)^3}\ln\frac{\xi+s}{1+s}+\frac{1}{\xi^4(\xi-1)^2(\xi+s)}
\end{equation*}
\begin{equation*}
-\,\frac{\xi(10\xi^2-15\xi+4)}{(\xi-1)^2(1+s)}-\frac{630\xi^5+175\xi^4+126\xi^3+63\xi^2+14\xi-9}{21\xi^4s}+80\xi^2\Hcsi(\xi,s)
\end{equation*}
\begin{equation}
+\,\frac{6(350\xi^4+105\xi^3+63\xi^2+28\xi+6)s^4-2\xi(630\xi^3+189\xi^2+98\xi+33)s^3+3\xi^2(252\xi^2+70\xi+27)s^2-30\xi^3(14\xi+3)s+180\xi^4}{126\xi^3s^6}.
\end{equation}

\subsubsection{The case $\xi=1$}
\begin{equation}
D(s)=-\,\frac{(6s^2 + 6s -1)(2s +1)}{2s^2 (1+s)^2}  - 6\ln\frac{s}{1+s},
\end{equation}
\begin{equation}
E(s)=-\,\frac{12s^4 + 30s^3 + 22s^2 + 3s + 3}{6s^2 (1+s)^3} - 2\ln\frac{s}{1+s},
\end{equation}
\begin{equation}
F(s)=\frac{60s^6 + 150s^5 + 110s^4 + 15s^3 - 3s^2 + s + 3}{6s^4(1+s)^3}+10\ln\frac{s}{1+s},
\end{equation}
\begin{equation*}
G(s)=\left[24\ln\frac{s}{1+s}+\frac{120s^5+60s^4-20s^3+10s^2-6s+4}{5s^5(1+s)}\right]\!\ln(1+s)-\frac{24}{5}\ln\frac{s}{1+s}
\end{equation*}
\begin{equation}
+\,\frac{576s^6+1260s^5+716s^4+9s^3-9s^2-24}{30s^4(1+s)^3}-24\Hcsi(1,s),
\end{equation}
\begin{equation*}
H(s)=\left[80\ln\frac{1+s}{s}-\frac{1680s^7+840s^6-280s^5+140s^4-84s^3+56s^2-40s+30}{21s^7(1+s)}\right]\!\ln(1+s)-\frac{18}{7}\ln\frac{s}{1+s}
\end{equation*}
\begin{equation}
-\,\frac{10404s^8+23490s^7+14314s^6+711s^5-243s^4+109s^3-57s^2-30s-180}{126s^6(1+s)^3}+80\Hcsi(1,s).
\end{equation}

\subsection{J3e models}

The functions $A$, $B$ and $C$ are the same as for J3e models. The functions
from $D$ to $H$ are instead given by
\begin{equation}
D(s)=A(s)\ln\frac{\xi+s}{\xi}+\frac{9\xi^2+3\xi+1}{3\xi^3}\ln\frac{\xi+s}{s}+\frac{1}{\xi-1}\ln\frac{\xi+s}{1+s}-\frac{2(1+3\xi)s-\xi}{6\xi^2 s^2}-4\Hcsi(\xi,s),
\end{equation}
\begin{equation}
E(s)=B(s)\ln\frac{\xi+s}{\xi}+\frac{9\xi^2-1}{3\xi^3}\ln\frac{\xi+s}{s}+\frac{5\xi-6}{(\xi-1)^2}\ln\frac{\xi+s}{1+s}+\frac{3\xi^2+\xi-1}{3\xi^2(\xi-1)(1+s)}-\frac{(\xi-2)s+\xi}{6\xi^2s^2(1+s)}-8\Hcsi(\xi,s),
\end{equation}
\begin{equation*}
F(s)=C(s)\ln\frac{\xi+s}{\xi}-\frac{100\xi^4+20\xi^3-5\xi-4}{10\xi^5}\ln\frac{\xi+s}{s}-\frac{8\xi-9}{(\xi-1)^2}\ln\frac{\xi+s}{1+s}+\frac{10\xi^4-20\xi^3-5\xi^2+\xi+4}{10\xi^4(\xi-1)(1+s)}
\end{equation*}
\begin{equation}
+\,\frac{3(40\xi^3+5\xi^2-6\xi-8)s^3\!+\xi(5\xi+4)(3-2\xi)s^2\!-4\xi^2(\xi+2)s\!+6\xi^3}{60\xi^4s^4(1+s)}+18\Hcsi(\xi,s),
\end{equation}
\begin{equation*}
G(s)=\left[12\ln\frac{1+s}{s}-\frac{60s^5+30s^4-10s^3+5s^2-3s+2}{5s^5(1+s)}\right]\!\xi^2\ln\frac{\xi+s}{\xi}+\frac{5\xi^2+5\xi+3}{5\xi^3}\ln\frac{\xi+s}{s}-\frac{\xi(12\xi^2-18\xi+5)}{(\xi-1)^2}\ln\frac{\xi+s}{1+s}
\end{equation*}
\begin{equation}
-\,\frac{30\xi^4-20\xi^3-2\xi-3}{5\xi^2(\xi-1)(1+s)}-\frac{(20\xi^3+5\xi^2+7\xi+6)s^3-\xi(10\xi^2+3\xi+3)s^2+2\xi^2(3\xi+1)s-4\xi^3}{10\xi^2s^4(1+s)}+12\xi^2\Hcsi(\xi,s),
\end{equation}
\begin{equation*}
H(s)=\left[40\ln\frac{s}{1+s}+\frac{840s^7+420s^6-140s^5+70s^4-42s^3+28s^2-20s+15}{21s^7(1+s)}\right]\!\xi^2\ln\frac{\xi+s}{\xi}+\frac{70\xi^4+21\xi^3-7\xi-6}{21\xi^5}\ln\frac{\xi+s}{s}
\end{equation*}
\begin{equation*}
+\,\frac{40\xi^3-60\xi^2+10\xi+9}{(\xi-1)^2}\ln\frac{\xi+s}{1+s}+\frac{420\xi^6-280\xi^5-140\xi^4+21\xi^3+7\xi^2-\xi-6}{21\xi^4(\xi-1)(1+s)}
\end{equation*}
\begin{equation*}
+\,\frac{280\xi^5+70\xi^4-42\xi^3-7\xi^2+8\xi+12}{42\xi^4s(1+s)}-40\xi^2\Hcsi(\xi,s)
\end{equation*}
\begin{equation}
-\,\frac{(420\xi^4+126\xi^3-14\xi^2+9\xi+18)s^4-\xi(252\xi^3+84\xi^2+5\xi+12)s^3+3\xi^2(56\xi^2+20\xi+3)s^2-15\xi^3(8\xi+3)s+90\xi^4}{126\xi^3s^6(1+s)}.
\end{equation}

\subsubsection{The case $\xi=1$}
\begin{equation}
D(s)=A(s)\ln(1+s)+\frac{13}{3}\ln\frac{1+s}{s}-\frac{2s^2+7s-1}{6s^2(1+s)}-4\Hcsi(1,s),
\end{equation}
\begin{equation}
E(s)=B(s)\ln(1+s)+\frac{8}{3}\ln\frac{1+s}{s}+\frac{32s^3+36s^2-1}{6s^2(1+s)^2}-8\Hcsi(1,s),
\end{equation}
\begin{equation}
F(s)=C(s)\ln(1+s)-\frac{111}{10}\ln\frac{1+s}{s}-\frac{138s^5+117s^4-34s^3+s^2+2s-2}{20s^4(1+s)^2}+18\Hcsi(1,s),
\end{equation}
\begin{equation*}
G(s)=\left[12\ln\frac{1+s}{s}-\frac{60s^5+30s^4-10s^3+5s^2-3s+2}{5s^5(1+s)}\right]\!\ln(1+s)+\frac{13}{5}\ln\frac{1+s}{s}
\end{equation*}
\begin{equation}
-\,\frac{146s^5+189s^4+22s^3-8s^2+4s-4}{10s^4(1+s)^2}+12\Hcsi(1,s),
\end{equation}
\begin{equation*}
H(s)=\left[40\ln\frac{s}{1+s}+\frac{840s^7+420s^6-140s^5+70s^4-42s^3+28s^2-20s+15}{21s^7(1+s)}\right]\!\ln(1+s)+\frac{26}{7}\ln\frac{1+s}{s}
\end{equation*}
\begin{equation}
+\,\frac{4572s^7+5598s^6+404s^5-206s^4+116s^3-72s^2+75s-90}{126s^6(1+s)^2}-40\Hcsi(1,s).
\end{equation}

\subsection{Functions needed for the projected velocity dispersion}\label{app:sigmalos_edgeon}

The asymptotic behaviour of $\siglos^2$ in equations \eqref{eq:a&b(lambda)} and \eqref{eq:c(lambda)} 
near the center, at the linear order in the flattenings, is determined by the functions  
\begin{equation}
\ag(\tX,\tY)=
\upi\,\frac{6\hspace{0.2mm}\Rtil^2\hspace{-0.4mm}+\hspace{0.1mm}3\hspace{0.2mm}\etas(2\tY^2\hspace{-0.4mm}-\tX^2)+\etag(3\tX^2\hspace{-0.4mm}+2\tY^2)}{12\hspace{0.2mm}\xi\Rtil^2}-k^2\hspace{0.4mm}\frac{4(3\hspace{0.2mm}\etas-\etag)\tY^2}{3\hspace{0.1mm}\xi\upi\Rtil^2},
\end{equation}
\begin{equation}
\aBH(\tX,\tY)=\frac{10\hspace{0.2mm}\Rtil^2\hspace{-0.4mm}+\hspace{0.1mm}6\hspace{0.2mm}\etas(3\tY^2\hspace{-0.4mm}-\tX^2)}{15\Rtil^2}-k^2 B^2\hspace{-0.8mm}\left(\frac{1}{2},\frac{5}{4}\right)\!\frac{4\hspace{0.2mm}\etas \tY^2}{5\upi\Rtil^2},
\end{equation}
while at large radii
\begin{equation}
b(\tX,\tY) = \frac{7\hspace{0.1mm}\Rtil^2\hspace{-0.4mm}+\hspace{0.1mm}\etas(31\hspace{0.1mm}\tY^2\hspace{-0.4mm}-17\tX^2)}{15\hspace{0.1mm}\Rtil^2}-k^2 B^2\hspace{-0.8mm}\left(\frac{1}{2},\frac{9}{4}\right)\!\frac{4\hspace{0.2mm}\etas \tY^2}{\upi\Rtil^2}.
\end{equation}

\end{document}